\documentclass[journal]{IEEEtran}
\usepackage{tikz}
\usetikzlibrary{arrows,positioning,fit,shapes,calc,decorations.pathreplacing,matrix,patterns}

\usepackage{amsmath}
\usepackage{subfigure}
\usepackage{lineno,hyperref}
\usepackage{xcolor}
\usepackage{extarrows}
\usepackage{algorithm}
\usepackage{algorithmic}
\usepackage{amssymb}
\usepackage{graphicx}
\usepackage[justification=centering]{caption}
\usepackage{cite}

\usepackage{stfloats}
\usepackage{amsthm}

\newcommand{\ba}{\boldsymbol{a}}

\newcommand{\bx}{\boldsymbol{x}}
\newcommand{\by}{\boldsymbol{y}}
\newcommand{\bh}{\boldsymbol{h}}
\newcommand{\bw}{\boldsymbol{w}}

\newcommand{\bp}{\boldsymbol{p}}
\newcommand{\bs}{\boldsymbol{s}}
\newcommand{\bq}{\boldsymbol{q}}
\newcommand{\br}{\boldsymbol{r}}

\newcommand{\be}{\boldsymbol{e}}

\newcommand{\bX}{\boldsymbol{X}}
\newcommand{\bE}{\boldsymbol{E}}
\newcommand{\bP}{\boldsymbol{P}}
\newcommand{\bY}{\boldsymbol{Y}}
\newcommand{\bA}{\boldsymbol{A}}
\newcommand{\bB}{\boldsymbol{B}}
\newcommand{\bR}{\boldsymbol{R}}
\newcommand{\bZ}{\boldsymbol{Z}}
\newcommand{\bW}{\boldsymbol{W}}
\newcommand{\bI}{\boldsymbol{I}}

\newcommand{\bC}{\boldsymbol{C}}
\newcommand{\bD}{\boldsymbol{D}}
\newcommand{\bU}{\boldsymbol{U}}
\newcommand{\bV}{\boldsymbol{V}}
\newcommand{\bQ}{\boldsymbol{Q}}
\newcommand{\bH}{\boldsymbol{H}}
\newcommand{\bS}{\boldsymbol{S}}
\newcommand{\bG}{\boldsymbol{G}}
\newcommand{\bJ}{\boldsymbol{J}}

\newcommand{\bXi}{\boldsymbol{\Xi}}
\newcommand{\bGamma}{\boldsymbol{\Gamma}}
\newcommand{\bLambda}{\boldsymbol{\Lambda}}

\newcommand{\N}{\mathcal{N}}

\newcommand{\mR}{\mathcal{R}}

\newcommand{\Tr}{\text{Tr}}
\newcommand{\tT}{\textrm{T}}
\newcommand{\tVec}{\textrm{Vec}}

\newcommand{\tra}{^{\textrm{T}}}

\newcommand{\MN}{\mathcal{MN}}

\newcommand{\bPhi}{\boldsymbol{\Phi}}

\begin{document}
\title{Unitary Approximate Message Passing for Matrix Factorization}
\author{Zhengdao~Yuan, Qinghua Guo, \IEEEmembership{Senior Member, IEEE}, Yonina C. Eldar, \IEEEmembership{Fellow, IEEE}, and Yonghui Li,   \IEEEmembership{Fellow, IEEE}
	\thanks{Corresponding author: Qinghua Guo.}
	\thanks{Z. Yuan is with the Artificial Intelligence Technology Engineering Research Center, Open University of Henan, 
	Zhengzhou 450002, China. He was with the School of Electrical, Computer and Telecommunications Engineering, University of Wollongong, Wollongong, NSW 2522, Australia (e-mail: yuan\_zhengdao@foxmail.com).}
	\thanks{Q. Guo is with the School of Electrical, Computer and Telecommunications Engineering, University of Wollongong, Wollongong, NSW 2522, Australia  (e-mail: qguo@uow.edu.au).}
	\thanks{Y. C. Eldar is with the Faculty of Mathematics and Computer Science, Weizmann Institute of Science, Rehovot 7610001, Israel (e-mail: yonina.eldar@weizmann.ac.il).}
	\thanks{Y. Li is with the School of Electrical and Information Engineering, University of Sydney, Sydney, NSW 2006, Australia (e-mail: yonghui.li@sydney.edu.au).}
}
\markboth{Unitary Approximate Message Passing for Matrix Factorization}
{Shell \MakeLowercase{\textit{et al.}}: Bare Demo of IEEEtran.cls for IEEE Journals}

\maketitle

\begin{abstract}
We consider matrix factorization (MF) with certain constraints, which finds wide applications in various areas.   
Leveraging variational inference (VI) and unitary approximate message passing (UAMP), we develop a Bayesian approach to MF with an efficient message passing implementation, called UAMP-MF. 
With proper priors imposed on the factor matrices, UAMP-MF can be used to solve many problems that can be formulated as MF, such as non-negative matrix factorization, dictionary learning, compressive sensing with matrix uncertainty, robust principal component analysis, and sparse matrix factorization. Extensive numerical examples are provided to show that UAMP-MF significantly outperforms state-of-the-art algorithms in terms of recovery accuracy, robustness and computational complexity. 
\end{abstract}


\section{Introduction}

We consider the problem of matrix factorization (MF) with the following model
\begin{align}
\bY=\bH\bX+\bW,  \label{eq:bilinear}
\end{align}
where $\bY\in\mR^{M\times L}$ denotes a known (observation) matrix, $\bW\in\mR^{M\times L}$ accounts for unknown perturbations or measurement errors,
and the matrices $\bH\in\mR^{M\times N}$ and $\bX\in\mR^{N\times L}$ are two factor matrices to be obtained. 
Depending on concrete application scenarios, the MF problem in \eqref{eq:bilinear} often has specific requirements on the matrices $\bH$ and $\bX$. For example, in non-negative matrix factorization (NMF) \cite{Lee1999Nature}, the entries of $\bH$ and $\bX$ need to take non-negative values. In dictionary learning (DL) \cite{Rubinstein2010DL}, matrix $\bX$ is sparse matrix and $\bH$ is a dictionary matrix to be learned. In compressive sensing with matrix uncertainty (CSMU) \cite{Zhu2011CSMU}, $\bX$ is sparse and the sensing matrix $\bH$ can be modeled as $\bH=\bar \bH +\bH'$, where the matrix $\bar \bH$ is known, and $\bH'$ denotes an unknown perturbation matrix. The robust principal component analysis (RPCA) problem \cite{Cand2011} can also be formulated as \eqref{eq:bilinear}, where both $\bH$ and $\bX$ admit specific structures \cite{Parker2014I}. We may also be interested in sparse MF, where both matrices $\bH$ and $\bX$ are sparse.

Many algorithms have been developed to solve the above problems.  For NMF, the multiplication (Mult) update algorithm proposed in \cite{Lee2001MIT} is often used, which is simple to implement, and produces good results. The alternating least squares (ALS) algorithm was proposed in  \cite{Berry2007} for NMF, where two least squares steps are performed alternately. The above two algorithms are very popular and are built in the MATLAB library. The projected gradient descent (AlsPGrad) \cite{Lin2007} algorithm, which is an improvement to Mult, usually has faster convergence speed. To solve the DL problem, one can use algorithms such as K-SVD (singular value decomposition) \cite{Aharon2006} and SPAMS (sparse modeling software) \cite{Mairal2010}. K-SVD is often effective, but can be trapped in local minima or even saddle points. 
The SPAMS algorithm often does not perform well when the number of training samples is small. For CSMU, the sparsity-cognizant total least squares (S-TLS) method was developed in \cite{Zhu2011Sparsity} by extending the classical TLS approach.
For the RPCA problem, many algorithms have been proposed. {The work in \cite{Cand2011} developed} iterative thresholding algorithms with low complexity, but their convergence is usually slow. The inexact augmented Lagrange multiplier algorithm (IALM) \cite{Lin2010} and low-rank matrix fitting algorithm (LMaFit) \cite{Wen2012} deliver better convergence speed and performance. 


Another line of techniques is based on the Bayesian framework, especially with the approximate message passing (AMP) algorithm \cite{Donoho2010a}, \cite{Donoho2010b} and its extension or variants. 
With Bayesian approaches, the requirements on matrices $\bH$ and $\bX$ for the factorization in \eqref{eq:bilinear} are incorporated by imposing proper priors on them. 
The AMP algorithm is employed due to its low complexity. AMP was originally developed for compressive sensing based on loopy belief propagation (BP) \cite{Donoho2010a}, 
and was later extended in \cite{rangan2011generalized} to solve the estimation problem with a generalized linear observation model. The bilinear generalized AMP (Bi-GAMP) algorithm was proposed in \cite{Parker2014I} to deal with the problem in \eqref{eq:bilinear}. However, Bi-GAMP inherits the shortcoming of the AMP algorithm, i.e., it is not robust to generic matrices $\bH$ and $\bX$. To achieve better robustness, 
a bilinear adaptive VAMP (BAd-VAMP) algorithm was proposed in \cite{Sarkar2019} and generalized in \cite{mengzhu}. However, for the general problem in \eqref{eq:bilinear}, to reduce computational complexity, BAd-VAMP does not allow the use of priors for matrix $\bH$, i.e., it is treated as a deterministic unknown matrix. This makes BAd-VAMP incapable of dealing with the requirements on $\bH$, e.g., RPCA or sparse MF where $\bH$ should be sparse. In addition, there is still considerable room for improvement in terms of performance, robustness and computational complexity. In \cite{yuan2021BiUTAMP}, with UAMP \cite{guo2015approximate} (which was formally called UTAMP), a more robust and faster approximate inference algorithm (called Bi-UTAMP in \cite{yuan2021BiUTAMP}) was proposed, which is designed for specially structured $\bH$ in \eqref{eq:bilinear}, i.e.,
$\bH=\sum_{k=1}^K b_k\bA_k$, where matrices $\{\bA_k\}$ are known, and coefficients $\{b_k\}$ (and matrix $\bX$) need to be estimated. 


{In this work, with the framework of variational inference (VI) \cite{winn2005variational}, we develop a more efficient Bayesian algorithm leveraging UAMP. The use of VI requires the updates of two vairational distributions on $\bX$ and $\bH$, which may be difficult and expensive 
With matrix normal distributions \cite{Waal1985} and through a covariance matrix whitening process, we incorporate UAMP into VI to deal with the updates of the variational distributions. The VI-based method is implemented using message passing with UAMP as its key component, leading to an algorithm called UAMP-MF.  
Inheriting the low complexity and robustness of UAMP, UAMP-MF is highly efficient and robust. It has cubic complexity per iteration, which is considered low for an MF problem (note that even the multiplication of two matrices has cubic complexity). Enjoying the flexibility of a Bayesian approach, UAMP-MF is able to handle various problems in a unified way with different priors.} We focus on its applications to NMF, DL, CSMU, RPCA and sparse MF, and compare its performance with various algorithms, including IAML, LMaFit and Bi-GAMP for RPCA, K-SVD, SPAMS and BAd-VAMP for DL, BAd-VAMP for the CSMU, AlsPGrad, and Mult for NMF. Extensive numerical results are provided to demonstrate the superiority of UAMP-MF in terms of recovery performance, robustness and computational complexity.


The remainder of this paper is organized as follows. In Section II, VI, (U)AMP and matrix normal distribution are briefly introduced. The UAMP-MF algorithm is derived in Section III. Applications of UAMP-MF to NMF, RPCA, DL, CSMU and sparse MF are investigated in Section IV, followed by conclusions in Section V.

{Throughout the paper, we use the following notations. Boldface lower-case and upper-case letters denote vectors and matrices, respectively, and superscript $(\cdot)^T$ represents the transpose operation.
A Gaussian distribution of $x$ with mean {$\hat x$} and variance $\nu_x$ is represented by $\N(x;{\hat x},\nu_x)$. 
Notation 
$\Tr(\cdot)$ denotes the trace operation.   
The relation $f(x)=cg(x)$ for some positive constant $c$ is written as $f(x)\propto g(x)$. The operation $diag(\ba)$ returns a diagonal matrix with $\ba$ on its diagonal. We use $\bA\cdot\bB$ and $\bA\cdot/\bB$ to represent the element-wise product and division between $\bA$ and $\bB$, respectively. 
We use $|\bA|^{.2}$ to denote element-wise magnitude squared operation for $\bA$, and $||\bA||$ is the Frobenius norm of $\bA$. 
The notations $\textbf{1}$, $\textbf{0}$ and $\bI$ denote an all-one matrix, an all-zero matrix and an identity matrix with a proper size, respectively. The above operations defined for matrices can also be applied to vectors. We use $\tVec(\cdot)$ to denote the vectorization operation.   

\section{Preliminaries} 

\subsection{Variational Inference}

VI is a machine learning method widely used to approximate posterior densities for Bayesian models \cite{winn2005variational}, \cite{jordan1999introduction}. Let $\bV$ and $\bR$ be the set of hidden (latent) variables and visible (observed) variables, respectively, with joint distribution $p(\bV,\bR)$. The goal of VI is to find a tractable variational distribution ${q}(\bV)$ that approximates the true posterior distribution $p(\bV|\bR)$. With the distribution $q(\bV)$, the log marginal distribution of the observed variables admits the following decomposition
\begin{equation}
	\ln p(\bR)=\mathcal{L}(q(\bV)) + \mathcal{KL}(q(\bV)||p(\bV|\bR)),
\end{equation}
where the variational lower bound $\mathcal{L}(q(\bV))$ is given as
\begin{equation}
	\mathcal{L}(q(\bV))=\int_{\bV} q(\bV) \ln \frac{p(\bV, \bR)}{q(\bV)},
\end{equation}
and the Kullback-Leibler (KL) divergence between $q(\bV)$ and $p(\bV|\bR)$ is
\begin{equation}
	\mathcal{KL}(q(\bV)||p(\bV|\bR))=-\int_{\bV} q(\bV) \log \frac{p(\bV| \bR)}{q(\bV)}.
\end{equation}
The distribution $q(\bV)$ that minimizes the KL divergence $\mathcal{KL}(q(\bV)||p(\bV|\bR))$ can be found by maximizing the variational lower bound $\mathcal{L}(q(\bV))$.

VI can be implemented with message passing with the assistance of graphical models such as factor graphs. If the variational distribution with some factorization is chosen, e.g.,
\begin{equation}
	q(\bV)=\prod_k q_k(\bV_k),
\end{equation}
where $\bV=\{\bV_k\}$,
then the variational distribution may be found through an iterative procedure \cite{winn2005variational} with the update rule
\begin{equation}
	q_k(\bV_k) \propto \exp \left(\int_{ \tilde {\bV}} q(\tilde{\bV}) \log f(\bV_k, \tilde{\bV})  \right).
\end{equation}
Here $ f(\bV_k, \tilde{\bV})$ is a local factor associated with $\bV_k$ and $\tilde{\bV} \in \{\bV_i, i \neq k \} $,
depending on the structure of the factor graph. The update of $\{q_k(\bV_k)\}$ is carried out iteratively until it converges or a pre-set number of iterations is reached.

\subsection{(U)AMP}

AMP was derived based on loopy BP with Gaussian and Taylor-series approximations  \cite{donoho2009message}, {\cite{rangan2011generalized}}, which can be used to recover $\bx$ from the noisy measurement $\by=\bA\bx+\bw$ with $\bw$ being a zero-mean white Gaussian noise vector. AMP works well with an i.i.d. Gaussian $\bA$, but it can easily diverge for generic $\mathbf{A}$ \cite{rangan2019convergence}.
Inspired by \cite{guo2013}, the work in \cite{guo2015approximate} showed that the robustness of AMP can be remarkably improved through simple pre-processing, i.e., performing a unitary transformation to the original linear model  \cite{yuan2021BiUTAMP}.
As any matrix $\bA$ has an SVD  $\bA= \bU \boldsymbol{\Lambda} \bV$ with $\bU$ and $\bV$ being two unitary matrices, performing a unitary transformation with $\bU\tra$ leads to the following model
\begin{equation}
\br=\mathbf{\Phi} \bx+\boldsymbol{\omega},
\label{r=uy}
\end{equation}
where $\br=\bU\tra \by$, $\boldsymbol{\Phi}=\bU\tra\bA=\boldsymbol{\Lambda}\bV$,
$\mathbf{\Lambda}$ is an $M\times N$ rectangular diagonal matrix, and $\boldsymbol{\omega} = \bU\tra \bw$ remains a zero-mean white Gaussian noise vector with the same covariance as $\boldsymbol{\omega}$. 

Applying the vector step size AMP {\cite{rangan2011generalized}} with model \eqref{r=uy} leads to the first version of UAMP {(called UAMPv1)} shown in Algorithm \ref{UTAMPv1}. Note that by replacing $\br$ and $\mathbf{\Phi}$ with $\by$ and $\bA$ in Algorithm \ref{UTAMPv1}, the original AMP algorithm is recovered. An average operation can be applied to two vectors: $\boldsymbol{\tau}_x$ in Line 7 and $|\mathbf{\Phi}^H |^{.2}  \boldsymbol{\tau}_s$ in Line 5 of {UAMPv1 in} Algorithm \ref{UTAMPv1}, leading to the second version of UAMP \cite{guo2015approximate} {(called UAMPv2), where the operations in the brackets of Lines 1, 5 and 7 are executed} (refer to \cite{yuan2021BiUTAMP} for the detailed derivations).
Compared to AMP and UAMPv1,  UAMPv2 does not require matrix-vector multiplications in Lines 1 and 5, so that the number of matrix-vector products is reduced from 4 to 2 per iteration. 

\begin{algorithm}
	\caption{UAMP (UAMPv2 executes operations in [ ])}
	Initialize $\boldsymbol{\tau}_x^{(0)} (\mathrm{or}~{\tau}_x^{(0)})>0$ and ${{\bx}^{(0)} }$. Set $\bs^{(-1)}=\mathbf{ 0 }$ and $t=0$. Define vector $\boldsymbol{\lambda}={\mathbf{ \Lambda \Lambda}\tra} \textbf{1}$.\\
	\textbf{Repeat}
	\begin{algorithmic}[1]
		\STATE $\boldsymbol{\tau}_p$ = $   \mathbf{|\Phi|}^{.2} \boldsymbol{\tau}^t_x$~~~~~~~~~~~~ $\left[\mathrm{or}~ \boldsymbol{\tau}_p =  \tau^t_x  \boldsymbol{\lambda}\right]$ \\
		\STATE $ \bp= \mathbf{\Phi}  {{\bx}^{t} } - \boldsymbol{\tau}_{p} \cdot  \bs^{t-1} $\\
		\STATE $ \boldsymbol{\tau}_s = \mathbf{1}./ (\boldsymbol{\tau}_p+\beta^{-1} \mathbf{1}) $\\
		\STATE $ \bs^t= \boldsymbol{\tau}_s \cdot (\br-\bp) $\\
		\STATE $\mathbf{1}./ \boldsymbol{\tau}_q$ = $ |\mathbf{\Phi}\tra |^{.2}  \boldsymbol{\tau}_s$~~~~~~~$\left[\mathrm{or}~ \mathbf{1}./ \boldsymbol{\tau}_q = (\frac{1}{N} \boldsymbol{\lambda}\tra \boldsymbol{\tau }_s) \mathbf{1}\right]$ \\
		\STATE $ \bq = {{\bx}^{t} } + \boldsymbol{\tau}_q \cdot(  \mathbf{\Phi}^H \bs^t)$\\
		\STATE $\boldsymbol{\tau}_x^{t+1}$ = $\boldsymbol{\tau}_q \cdot g_{x}' ( \bq, \boldsymbol{\tau}_q)$~~~~~~$\left[\mathrm{or}~ \tau_x^{t+1} \!=\!  \frac{1}{N}  \mathbf{1}\tra  \left(\boldsymbol{\tau}_q \cdot g_{x}' ( \bq, \tau_q)\right) \right]$ \\
		\STATE 	$ {{\mathbf{x}}^{t+1} } = g_{x}  ( \bq, \boldsymbol{\tau}_q)$\\	
		\STATE 	$  t=t+1$
	\end{algorithmic}
	\textbf{Until terminated}
	\label{UTAMPv1}
\end{algorithm}


In the (U)AMP algorithms, $g_x(\bq, \boldsymbol{\tau}_q )$ is related to the prior of $\bx$ and returns a column vector with $n$th
entry $[ g_x(\mathbf{q}, \boldsymbol{\tau}_q ) ]_n$ given by
\begin{equation}
[g_x(\bq, \boldsymbol{\tau}_q ) ]_n
=
\frac{\int x_n p(x_n) \mathcal{N} (x_n ; q_n, \tau_{q_n})  d x_n }{\int  p(x_n) \mathcal{N} (x_n ; q_n, \tau_{q_n})  d x_n },
\label{g_x}
\end{equation}
where  $p(x_n)$ represents a prior for $x_n$.
The function $g_x'(\bq,\boldsymbol{\tau}_q)$ returns a column vector and the $n$th element is denoted by $[ g_x'(\bq, \boldsymbol{\tau}_q ) ]_n$, where the derivative is taken with respect to $q_n$.
	
\subsection{Matrix Normal Distribution}

{In this work, the matrix normal distribution is used in the derivation of VI for the MF problem, into which UAMP is incorporated, leading to the UAMP-MF algorithm. The matrix normal distribution is a generalization of the multivariate normal distribution to matrix-valued random variables \cite{Waal1985}. The distribution of normal random matrix $\bX$ is
\begin{eqnarray}
	&&\!\!\!\!\!\!\!\!\!\!\MN\big(\bX;\hat\bX,\bU_X,\bV_X\big)=(2 \pi)^{-\frac{NL}{2}}
	\big|\bU_X\big|^{-\frac{N}{2}}\big|\bV_X\big|^{-\frac{L}{2}}\nonumber\\
	&&\exp\Big(-\frac{1}{2}\Tr\big( \bV_X^{-1}\big(\bX-\hat\bX\big)^\tT\bU_X^{-1}\big(\bX-\hat\bX\big)\big)\Big),
	\label{eq:MN_Distrib}
\end{eqnarray}
where $\hat\bX\in \mR^{N\times L}$ is the mean of $\bX$,  $\bU_X\in\mR^{N\times N}$ and $\bV_X\in\mR^{L\times L}$ are the covariance among rows and columns of $\bX$, respectively.
The matrix normal distribution is related to the multivariate normal distribution in the following way  
\begin{eqnarray}
	\bx \sim \N\big(\bx;\hat\bx,\bV_X\otimes \bU_X\big), 
\end{eqnarray}
where $\bx=\tVec(\bX)$ and $\hat\bx=\tVec(\hat\bX)$.}



\section{Design of UAMP-MF}

{The Bayesian treatment of the MF problem leads to great flexibility. Imposing proper priors $p(\bH)$ and $p(\bX)$ on $\bH$ and $\bX$, we can accommodate many concrete MF problems. 
For example in DL, we aim to find sparse coding of data set $\bY$, where $\bH$ is the dictionary and $\bX$ is the sparse representation with the dictionary. For the entries of $\bH$, we can use a Gaussian prior 
\begin{eqnarray}\label{adda}
	p(\bH)=\prod\nolimits_{m,n}p(h_{m,n})=\prod\nolimits_{m,n} \mathcal{N}(h_{m,n};0,1),
\end{eqnarray}
while for the entries of $\bX$, we can consider a sparsity promoting hierarchical Gaussian-Gamma prior
\begin{eqnarray} 
	p(\bX)&=&\prod\nolimits_{n,l}p(x_{n,l})=\prod\nolimits_{n,l} \mathcal{N}(x_{n,l};0,\gamma_{n,l}^{-1}) \label{adda1} \\
	p(\gamma_{n,l})&=&Ga(\gamma_{n,l};\epsilon,\eta). \label{add1a}
\end{eqnarray}
In Section IV,  more examples on the priors, including NMF, RPCA, CSMU and sparse MF, are provided.}  

{With the priors and model (1),  if the a posteriori distributions $p(\bH|\bY)$ and $p(\bX|\bY)$ can be found, then the estimates of $\bH$ and $\bX$ are obtained, e.g., using the a posteriori means of $\bH$ and $\bX$ to serve as their estimates. However, it is intractable to find the exact a posteriori distributions in general, so we resort to the approximate inference method VI to find their variational approximations $q(\bH) \approx p(\bH|\bY)$ and $q(\bX)\approx p(\bX|\bY)$. Finding the varational distributions are still challenging due to the large dimensions of $\bH$ and $\bX$, and the priors of $\bH$ and $\bX$ may also lead to intractable $q(\bH)$ and $q(\bX)$. In this work, through a covariance matrix (model noise) whitening process, UAMP is incorporated to solve these challenges with high efficiency. This also leads to Gaussian variational distributions $q(\bH)$ and $q(\bX)$, so that the estimates of $\bH$ and $\bX$ (i.e., the a posteriori means of $\bH$ and $\bX$) appear as the parameters of the variational distributions.} 

{Leveraging UAMP, we carry out VI in a message passing manner, with the aid of a factor graph representation of the problem. This leads to the message passing algorithm UAMP-MF. In the derivation of UAMP-MF, we use the notation $m_{n_a \rightarrow  n_b} (\bX)$ to denote a message passed from node $n_a$ to node $n_b$, which is a function of $\bX$. The algorithm UAMP-MF is shown in Algorithm \ref{alg1}, which we will frequently refer to in this section.}


\subsection{Factor Graph Representation}

\begin{algorithm}
	\caption{UAMP-MF}
	\textbf{Initialization}: $\bU_H=\bI_M$, $\bV_H=\bI_N$, $\hat\bH=\textbf{1}_{MN}$, $\bV_X=\bI_L$.
	$\bXi_X= \textbf{1}_{NL}$, $\bS_X=\textbf{0}_{NL}$,
	$\bXi_H=\textbf{1}_{MN}$, $\bS_H=\textbf{0}_{MN}$.
	\textbf{Repeat}
	\begin{algorithmic}[1]
		\STATE$\overline\bW_X=\hat\bH\tra\hat\bH+ M \bV_H$\\
		\STATE$[\bC_X,\bD_X]=\text{eig}(\overline\bW_X)$ \\
		\STATE$\bR_X = \bD_X^{-\frac{1}{2}}\bC_X\tra\hat\bH\tra\bY$,
		$\bPhi_X= \bD_X^{-\frac{1}{2}}\bC_X\tra$\\
		\STATE$\bV_{P_X}=|\bPhi_X|^{.2}\bXi_X$ \\
		\STATE $\bP=\bPhi_X\hat\bX-\bV_{\bP_X}\cdot\bS_X$\\
		\STATE$\bV_{S_X}=\textbf{1}./(\bV_{P_X}+\hat\lambda^{-1}\mathbf{1})$ \\
		\STATE$\bS_X=\bV_{S_X}\cdot(\bR_X-\bP_X)$\\
		\STATE$\bV_{Q_X}=\textbf{1}./(|\bPhi_X\tra|^{.2}\bV_{S_X})$\\
		\STATE$\bQ_X=\hat\bX+\bV_{Q_X}\cdot(\bPhi_X\tra\bS_X)$\\
		\STATE{$\bXi_X=\bV_{Q_X}\cdot\bG_X'(\bQ_X,\bV_{Q_X})$}\\
		\STATE$\hat\bX=\bG_X(\bQ_X,\bV_{Q_X})$\\
		\STATE$\bU_X=\text{diag}(\text{mean}(\bXi_X,2))$\\
		\STATE$\overline\bW_H=\hat\bX\hat\bX\tra+ L\bU_X$  \STATE$[\bC_H,\bD_H]=\text{eig}(\overline\bW_H)$ \\
		\STATE$\bR_H = \bD_H^{-\frac{1}{2}}\bC_H\tra\hat\bX\bY\tra$,
		$\bPhi_H = \bD_H^{-\frac{1}{2}}\bC_H\tra$\\
		\STATE$\bV_{P_H}=|\bPhi_H|^{.2}\bXi_H$\\ \STATE$\bP_H=\bPhi_H\hat\bH-\bV_{\bP_H}\cdot\bS_H$\\
		\STATE$\bV_{S_H}=\mathbf{1}./(\bV_{P_H}+\hat\lambda^{-1}\mathbf{1})$
		\STATE$\bS_H=\bV_{S_H}\cdot(\bR_H-\bP_H)$\\
		\STATE$\bV_{Q_H}=\mathbf{1}./(|\bPhi_H\tra|^{.2}\bV_{S_H})$\\
		\STATE$\bQ_H=\hat\bH+\bV_{Q_H}\cdot(\bPhi_H\tra\bS_H)$\\
		\STATE{$\bXi_H=\bV_{Q_H}\cdot\bG_H'(\bQ_H,\bV_{Q_H})$}\\
		\STATE $\hat\bH=\bG_H(\bQ_H,\bV_{Q_H})$\\
		\STATE$\bU_H=\text{diag}(\text{mean}(\bXi_H,1))$ \\
		\STATE$\hat\lambda=ML/C$ with $C$ given in \eqref{eq:ComputeC}
	\end{algorithmic}
	\textbf{Until terminated}
	\label{alg1}
\end{algorithm}

With model \eqref{eq:bilinear}, we have the following joint conditional distribution and its factorization
\begin{eqnarray}
	p(\bX,\bH,\lambda|\bY) \!\!\!\!\!&\propto&\!\!\!\!\!  p(\bY|\bX,\bH,\lambda)p(\bX)p(\bH)p(\lambda)\nonumber\\
	\!\!\!\!\!&\triangleq&\!\!\!\!\! f_{Y} (\bY, \! \bX, \!\bH, \!\lambda) f_{X}(\bX) f_{H} (\bH) f_{\lambda} (\lambda) \label{eq:factor}
\end{eqnarray}
where
\begin{eqnarray}
	f_{Y} (\bY, \bX, \bH, \lambda)&\triangleq&p\big(\bY|\bX,\bH,\lambda\big)\nonumber\\
	&=& \MN\big(\bY;\bH\bX,\bI_M,\lambda^{-1}\bI_L\big).
\end{eqnarray}
The precision $\lambda$ of the noise has an improper prior $f_{\lambda}(\lambda)\triangleq p(\lambda)\propto 1/\lambda$ \cite{Tipping}, and the priors for $\bH$ and $\bX$ denoted by $f_{H}(\bH)\triangleq p(\bH)$ and $f_{X}(\bX)\triangleq p(\bX)$ are customized according to the requirements in a specific problem. 
The factor graph representation of $\eqref{eq:factor}$ is depicted in Fig.\ref{fig:FactorGraph}, where the arguments of the local functions are dropped for the sake of conciseness. We define a variational distribution
\begin{equation}
	q(\bX, \bH, \lambda)=q(\bX)q(\bH)q(\lambda),
\end{equation}
{and we expect that $q(\bX) \approx p(\bX|\bY)$, $q(\bH)\approx p(\bH|\bY)$ and $q(\lambda) \approx p(\lambda|\bY)$}. 
Next, we study how to efficiently update $q(\bX)$, $q(\bH)$ and $q(\lambda)$ iteratively. 

\begin{figure}[!t]
	\centering
	\includegraphics[width=0.31\textwidth]{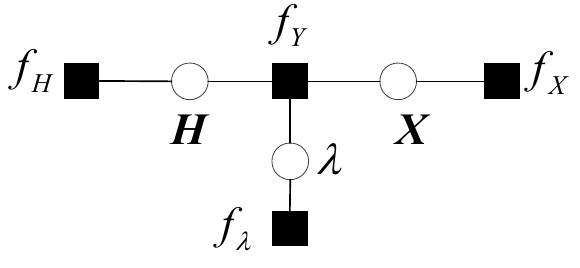}
	\centering
	\caption{Factor graph representation of $\eqref{eq:factor}$.}
	\label{fig:FactorGraph}
\end{figure}

\subsection{Update of $q(\bX)$}
According to VI, with $q(\bH)$ and $q(\lambda)$ (updated in the previous iteration), we update $q(\bX)$. As shown by the factor graph in Fig. \ref{fig:FactorGraph}, we need to compute the message $m_{f_Y\to X}(\bX)$ from the factor node $f_Y$ to the variable node $\bX$ and then combine it with the prior $p(\bX)$. Later we will see that $q(\bH)$ is a matrix normal distribution, i.e., $q(\bH)=\MN(\bH;\hat\bH,\bU_H,\bV_H)$ with $\bU_H=\bI_{N}$, and $q(\lambda)$ is a Gamma distribution. It turns out that the message $m_{f_Y\to X}(\bX)$ is matrix normal, as shown by Proposition 1.

\vspace{0.1 in}
\textbf{Proposition 1}: The message from $f_Y$ to $\bX$ is a matrix normal distribution of $\bX$, which can be expressed as
\begin{equation}
m_{f_Y\to X}(\bX)\propto \MN(\bX;\overline\bX, \hat\lambda^{-1}\overline\bU_X,\bI_L), \label{eq:msg_fyX_result}
\end{equation}
with
\begin{equation}
\overline\bX = \overline\bU_X\hat\bH\tra\bY, \label{eq:msg_fyX_mean} 
\end{equation}
\begin{equation}
\overline\bU_X = \big(\hat\bH\tra\hat\bH+\Tr(\bU_H)\bV_H\big)^{-1}, \label{eq:msg_fyX_U}
\end{equation}
and
\begin{equation}
\hat\lambda=\int_{\lambda} \lambda q(\lambda).
\end{equation}
The computation of $\hat \lambda$ is shown in \eqref{eq:lambda}.


\begin{proof}
See Appendix A.
\end{proof}

According to VI, this message is combined with the prior $p(\bX)$ to obtain $q(\bX)$. This can be challenging as $p(\bX)$ may lead to an intractable $q(\bX)$ and high computational complexity. Next, leveraging UAMP, we update $q(\bX)$ efficiently. It is also worth mentioning that, with the use of UAMP, the matrix inversion involved in \eqref{eq:msg_fyX_U} does not need to be performed. 

Note that $\bX=[\bx_1,...,\bx_l]$ and $\overline\bX=[\overline\bx_1,...,\overline\bx_l]$. The result in \eqref{eq:msg_fyX_result} indicates that $\bx_l \sim \N(\bx_l;\overline\bx_l, \hat\lambda^{-1}\overline\bU_X)$. This leads to the following pseudo observation model
\begin{eqnarray}
\overline\bx_l=\bx_l+\be_x, \label{eq:pseudoX1}
\end{eqnarray}
where $\be_x\sim \N(\be_x; 0,\hat\lambda^{-1}\overline\bU_X)$, i.e., the model noise is not white. We can whiten the noise by left-multiplying  
both sides of \eqref{eq:pseudoX1} by $\overline\bU_X^{-\frac{1}{2}}$, leading to
\begin{eqnarray}
\overline\bU_X^{-\frac{1}{2}}\overline\bx_l
= \overline\bU_X^{-\frac{1}{2}}\bx_l+\bw_x,
\end{eqnarray}
where $\bw_x=\overline\bU_X^{-\frac{1}{2}}\be_x$ is white and Gaussian with covariance $\hat\lambda^{-1}\bI$. Through the whitening process, we get a standard linear model with white additive Gaussian noise, which facilitates the use of UAMP. Considering all the vectors in $\overline\bX$, we have  
\begin{eqnarray}
\overline\bU_X^{-\frac{1}{2}}\overline\bX=\overline\bU_X^{-\frac{1}{2}}\bX+ \boldsymbol{\Omega_X}, \label{eq:pseudoX2}
\end{eqnarray}
where $\boldsymbol{\Omega_X}$ is white and Gaussian.

With model \eqref{eq:pseudoX2} and the prior $p(\bX)$, we use UAMP to update $q(\bX)$. Following UAMP, a unitary transformation needs to be performed with the unitary matrix $\bC_X^T$ obtained from the SVD of $\overline\bU_X^{-\frac{1}{2}}$ (or eigenvalue decomposition (EVD) as the matrix is definite and symmetric), i.e.,
\begin{eqnarray}
\overline\bU_X^{-\frac{1}{2}} = \bC_X \bLambda\bC_X^T
\end{eqnarray}
with $\bLambda$ being a diagonal matrix. After performing the unitary transformation, we have
\begin{eqnarray}
\bR_X=\bPhi_X \bX+ \boldsymbol{\Omega'_X},\label{eq:pseudoX3}
\end{eqnarray}
where $\bR_x=\bC_X^T\overline\bU_X^{-\frac{1}{2}}\overline\bX$, $\bPhi_X=\bC_X^T\overline\bU_X^{-\frac{1}{2}}=\bLambda\bC_X^T$  and $ \boldsymbol{\Omega'_X}=\bC_X^T \boldsymbol{\Omega_X}$, which is still white and Gaussian.

From the above, it seems costly to obtain $\bR_X$ and $\bPhi_X$ in model \eqref{eq:pseudoX3}. Specifically, a matrix inverse operation needs to be performed according to \eqref{eq:msg_fyX_U} so that $\overline \bX$ in \eqref{eq:msg_fyX_mean} can be obtained; a matrix squared root operation is needed to obtain $\overline\bU_X^{-\frac{1}{2}}$; and an SVD operation is required to $\bC_X^T$. Next, we show that the computations can be greatly simplified.

Instead of computing $\overline\bU_X^{-1}$ and $\overline\bU_X^{-\frac{1}{2}}$ followed by SVD of $\overline\bU_X^{-\frac{1}{2}}$, we perform EVD to $\overline\bU_X^{-1}=\overline\bW_X=\hat\bH\tra\hat\bH+\Tr(\bU_H)\bV_H$, i.e.,
\begin{eqnarray}
	[\bC_X,\bD_X]=\text{eig}(\overline\bW_X),
\end{eqnarray}
where the diagonal matrix
\begin{equation}
	\bD_X= \bLambda^{-2}.
\end{equation}
Hence $\bPhi_X=\bLambda\bC_X^T=\bD_X^{-1/2}\bC_X^T$, where the computation of $\bD_x^{-1/2}$ is trivial as $\bD_x$ is a diagonal matrix. Meanwhile, the computation of the pseudo observation matrix $\bR_x$ can also be simplified:
\begin{eqnarray}
\bR_X&=&\bC_X^T\overline\bU_X^{-\frac{1}{2}}\overline\bX \nonumber \\
&=&\bC_X^T\overline\bU_X^{-\frac{1}{2}}\overline\bU_X\hat\bH\tra\bY \nonumber\\
&=& {\bD_X^{-\frac{1}{2}}\bC_X\tra\hat\bH\tra\bY}.
\end{eqnarray}
For convenience, we rewrite the unitary transformed pseudo observation model as
\begin{eqnarray}
	{\underbrace{\bD_X^{-\frac{1}{2}}\bC_X\tra\hat\bH\tra\bY}_{\bR_X}}
	=\underbrace{\bD_X^{-\frac{1}{2}}\bC_X\tra}_{\bPhi_X}\bX+
	\boldsymbol{\Omega'_X}. \label{eq:pseudoX4}
\end{eqnarray}
The above leads to Lines 1-3 of UAMP-MF in Algorithm 2. 

Due to the prior $p(\bX)$, the use of exact $q(\bX)$ often makes the message update intractable. Following UAMP in Algorithm 1, we project it to be Gaussian, which can also be interpreted as performing the minimum mean squared estimation (MMSE) based on the pseudo observation model  \eqref{eq:pseudoX4} with prior $p(\bX)$. These correspond to Lines 4-11 of UAMP-MF in Algorithm 2. We assume that the prior $p(\bX)$ are separable, i.e., $p(\bX)=\prod_{n,l}p(x_{nl})$, then the operations in Lines 10 and 11 are element-wise as explained in the following. As the estimation is decoupled by (U)AMP, we assume the following scalar pseudo models
\begin{equation}
	q_{nl}=x_{nl}+w_{nl}, n=1,...,N, l=1,...,L, \label{eq:scalarmodel}
\end{equation}   
where $q_{nl}$ is the $(n,l)$th element of $\bQ_X$ in Line 9 of the UAMP-MF algorithm, $w_{nl}$ represents a Gaussian noise with mean zero and variance $v_{nl}$, and $v_{nl}$is the $(n,l)$th element of $\bV_{Q_x}$ in Line 8 of the UAMP-MF algorithm. This is significant as the complex estimation is simplified to much simpler MMSE estimation based on a number of scalar models \eqref{eq:scalarmodel} with prior $p(x_{nl})$. With the notations in Lines 10 and 11, for each entry $x_{nl}$ in $\bX$, the MMSE estimation leads to a Gaussian distribution 
\begin{eqnarray}
q'(x_{nl})=\N(x_{nl};\hat x_{nl},v_{x_{nl}}),
\end{eqnarray}
where $\hat x_{nl}$ and $v_{x_{nl}}$ is the $(n,l)$th element of $\hat\bX$ in Line 11 and the $(n,l)$th element of $\bXi_X$ in Line 10, respectively. We can see that each element $x_{nl}$ has its own variance. To facilitate subsequent processing, we make an approximation by assuming that the elements of each row in $\hat\bX$ have the same variance, which is the average of their variances. Then they are collectively characterized by a matrix normal distribution, i.e., $q(\bX)=\MN(\bX; \hat\bX, \bU_X, \bV_X)$ with $\bU_X=\text{diag}(\text{mean}(\bXi_X,2))$ and $\bV_X=\bI_L$, where  $\text{mean}(\bXi_X,2)$ represents the average operation on the rows of $\bXi_X$. This leads to Line 12 of UAMP-MF in algorithm 2. 
    


\subsection{Update of $q(\bH)$}
With the updated $q(\bX)=\MN(\bX; \hat\bX, \bU_X, \bV_X)$ and $q(\lambda)$, we compute the message from $f_Y$ to $\bH$ according to VI. Regarding the message, we have the following result.

\vspace{0.1 in}
\textbf{Proposition 2}: The message from $f_Y$ to $\bH$ is a matrix normal distribution about $\bH$, which can be expressed as
\begin{equation}
	m_{f_Y\to H}(\bH)\propto \MN(\bH;\overline\bH,\bI_M, \hat\lambda^{-1} \overline\bV_H), \label{eq:msg_fyH_result}
\end{equation}
with
\begin{equation}
	\overline\bH =\bY\hat\bX\tra\overline\bV_H. \label{eq:msg_fyH_mean}
\end{equation}
and
\begin{equation}
	\overline\bV_H = \left(\hat\bX\hat\bX\tra+\Tr(\bV_X)\bU_X\right)^{-1}, \label{eq:msg_fyH_V}
\end{equation}
where
$\hat \lambda=\int_{\lambda} \lambda q(\lambda)$. 

\begin{proof}
	See Appendix B.
\end{proof}

Then we combine the message $m_{f_Y\to H}(\bH)$ with the prior of $\bH$ to update $q(\bH)$. This will also be realized with UAMP through a whitening process. The procedure is similar to that for $q(\bX)$, and the difference is that the pseudo model is established row by row (rather than column by column) by considering the form of $m_{f_Y\to H}(\bH)$.

With the message $m_{f_Y\to H}(\bH)$ and
\begin{equation}
\overline\bH=
\left(
\begin{matrix}
\overline\bh_1\tra \\
\vdots \\
\overline\bh_M\tra
\end{matrix}
\right),
\end{equation}
where $\overline\bh_m \tra\in\mR^{1\times N}$ is  the $m$th row vector, we have the pseudo observation model
\begin{eqnarray}
\overline\bh_m=\bh_m+\be_h \label{eq:pseudoH1},
\end{eqnarray}
where $\be_h\sim\N(\be_h;0, \hat \lambda^{-1} \overline\bV_H)$.
Whitening operation to \eqref{eq:pseudoH1} leads to
\begin{eqnarray}
\overline\bV_H^{-\frac{1}{2}}\overline\bh_m =\overline\bV_H^{-\frac{1}{2}}\bh_m+ \bw_h ,
\end{eqnarray}
where $\bw_h = \overline\bV_H^{-\frac{1}{2}} \be_h $, which is white and Gaussian with covariance $\hat\lambda^{-1}\bI$.
Collecting all rows and representing them in matrix form, we have
\begin{eqnarray}
\overline\bV_H^{-\frac{1}{2}}\overline\bH\tra = \overline\bV_H^{-\frac{1}{2}}\bH\tra+ \boldsymbol{\Omega}_H. \label{eq:pseudoH2}
\end{eqnarray}
UAMP is then employed with the model \eqref{eq:pseudoH2}. Using the ideas in the previous section for updating $q(\bX)$, we obtain the unitary transformed model efficiently.

We first perform an EVD to matrix $\overline \bV_H^{-1}=\overline\bW_H=\hat\bX\hat\bX\tra+\Tr(\bV_X)\bU_X$, i.e.,
\begin{eqnarray}
[\bC_H,\bD_H]=\text{eig}(\overline\bW_H).
\end{eqnarray}
Then the unitary transformed model is given as
\begin{eqnarray}
\bR_H\tra=\bPhi_H\tra \bH\tra+\boldsymbol{\Omega}'_H
\end{eqnarray}
where
\begin{eqnarray}
\bPhi_H &=& \bD_H^{-\frac{1}{2}}\bC_H\tra \\
\bR_H &=& \bD_H^{-\frac{1}{2}}\bC_H\tra\hat\bX\bY\tra,
\end{eqnarray}
and $\boldsymbol{\Omega_H}$ is white and Gaussian. The above leads to Lines 13-15 of UAMP-MF in Algorithm 2. 

Following UAMP in Algorithm 1, we obtain $q(\bH)$ and project it to be Gaussian, which correspond to Lines 16-23 of UAMP-MF Algorithm 2. Similar to the case for updating $q(\bX)$, 
to accommodate $\{q(h_{mn})\}$ with a matrix normal distribution, we make an approximation by assuming the entries in each column of $\bH$ share a common variance, which is the average of their variances. Then $q(\bH)=\MN(\bH;\hat\bH,\bU_H,\bV_H)$ with $\bV_H=\text{diag}(\text{mean}(\bXi_H,1))$ and $\bU_H=\bI_M$ (i.e., Line 24 of UAMP-MF in Algorithm 2), where $\text{mean}(\bXi_H,1)$ represents the average operation on the columns of $\bXi_H$.    


\subsection{Update of $q(\lambda)$}
With $q(\bH)=\MN\left(\bH; \hat\bH, \bV_H, \bI_M \right)$ and $q(\bX)=\MN\left(\bX; \hat\bX, \bI_L, \bU_X\right)$, we update $q(\lambda)$, i.e., combining the message $m_{f_Y\to \lambda}(\lambda)$ from $f_Y$ to $\lambda$ and its prior $p(\lambda)$. 

\vspace{0.1 in}
\textbf{Proposition 3}: The message from $f_Y$ to $\lambda$ can be expressed as
\begin{eqnarray}
	m_{f_Y\to \lambda}(\lambda)\propto\lambda^{ML}\exp\Big(-\lambda C \Big), \label{eq:msg_fy_lambda}
\end{eqnarray}
where
\begin{eqnarray}
&& \!\!C=\| \bY-\hat\bH\hat\bX\|_F^2+M\Tr\Big(\hat\bX\hat\bX\tra\bV_H\Big)\nonumber\\
&& \ \ \ \ \ \ \ \ +L \Tr(\bU_X\hat\bH\tra\hat\bH)+ML\Tr(\bU_X \bV_H). \label{eq:ComputeC}
\end{eqnarray}

\begin{proof}
See Appendix C. 	
\end{proof}

With the message $m_{f_Y\to \lambda}(\lambda)$ and the prior $p(\lambda)\propto 1/\lambda$, $q(\lambda)$ can be computed as
\begin{equation}
	q(\lambda)\propto m_{f_Y\to \lambda}(\lambda)p(\lambda) = \lambda^{ML-1}\exp\Big(-\lambda C \Big),
\end{equation}
which is a Gamma distribution. 
Then the mean of $\lambda$ is obtained as
\begin{eqnarray}
\hat\lambda = {\int_{\lambda} \lambda q(\lambda )}
={ML}/{C}, \label{eq:lambda}
\end{eqnarray}
which is Line 25 of UAMP-MF.



\subsection{Discussion and Computational Complexity}

Regarding UAMP-MF in Algorithm 2, we have the following remarks:
\begin{itemize}
	\item We do not specify the priors of $\bH$ and $\bX$ in Algorithm 2. The detailed implementations of Lines 10, 11, 22 and 23 of the algorithm depends on the priors in a concrete scenario. Various examples are provided in Section IV.
	\item As the MF problem often has local minima, we can use the strategy of restart to mitigate the issue of being stuck at local minima. In addition, the iterative process can be terminated based on some criterion, e.g., the normalized difference between the estimates of two consecutive iterations is smaller than a threshold.
	\item There are often hyper-parameters in the priors of $\bH$ and $\bX$. If we do not have knowledge on these hyper-parameters, their values need to be leaned or tuned automatically. Thanks to the factor graph and message passing frame work, these extra tasks can be implemented by extending the factor graph in Fig. 1. 
\end{itemize}

%
%
%

\tikzstyle{factornode} = [draw, fill=white, circle, inner sep=1pt,minimum size=0.4cm]
\tikzstyle{funnode} = [draw, rectangle,fill=black!100, minimum size = 0.4cm]
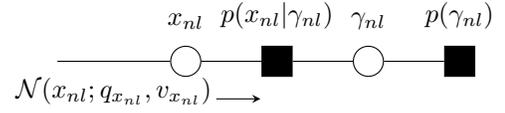
\begin{figure}[!t]
	\centering
	\begin{tikzpicture}[auto, node distance=0.8cm,>=latex']
		\node (X)[factornode,] at (0,0) {};
		\node (PXGamma)[funnode, right=0.8cm of X] {};
		\node (Gamma)[factornode, right=0.8cm of PXGamma] {};
		\node (PGamma)[funnode, right=0.8cm of Gamma] {};
		
		\node [above = 0.1cm of X] {$x_{nl}$};
		\node [above = 0.1cm of PXGamma] {$p(x_{nl}|\gamma_{nl})$};
		\node [above = 0.1cm of Gamma] {$\gamma_{nl}$};
		\node [above = 0.1cm of PGamma] {$p(\gamma_{nl})$};
		
		\draw ($(X.west)+(-1.5, 0)$) --node[below = 0.1cm of X]{$\mathcal{N}(x_{nl}; {q}_{x_{nl}}, v_{x_{nl}})$} (X.west);
		\draw (X) -- (PXGamma) -- (Gamma);
		\draw (Gamma) -- (PGamma);
		
		\draw [ ->, >=stealth] ($(X.east)+(0.2, -0.5)$) -- ($(X.west)+(1.2, -0.5)$);
	\end{tikzpicture}
	\centering
	\caption{Factor graph for learning hyper-parameters.}
	\label{fig:FactorGraph2}
\end{figure}

In the following, we show how to learn the hyper-parameters of the prior of $\bX$. We assume that $\bX$ is sparse and the sparsity rate is unknown. In this case, we may employ the sparsity promoting hierarchical Gaussian-Gamma prior \cite{Tipping}. The prior is then   $p(x_{nl},\gamma_{nl})= p(x_{nl}|\gamma_{nl})p(\gamma_{nl})$, where $p(x_{nl}|\gamma_{nl})=\N(x_{nl};0,\gamma_{nl}^{-1})\triangleq f_{x_{nl}}$ and $p(\gamma_{nl})=Ga(\gamma_{nl}; \epsilon, \eta)$ with $\epsilon$ and $\eta$ being the shape parameter and scale parameter of the Gamma distribution, respectively. While the precision $\gamma_{nl}$ is to be learned, the values for $\epsilon$ and $\eta$ are often set empirically, e.g., $\epsilon= \eta=0$ \cite{Tipping}. It is worth mentioning that $\epsilon$ can be tuned automatically to improve the performance \cite{UAMPSBL}. 

The factor graph representation for this part is shown in Fig. 2. Also note that UAMP decouples the estimation as shown by the pseudo model \eqref{eq:scalarmodel}, indicating that the incoming message to the factor graph in Fig. 2 is Gaussian, i.e., 
\begin{equation}
m_{x_{nl}\to f_{x_{nl}}}(x_{nl})=\mathcal{N} (x_{nl}; q_{nl}, v_{nl}). 
\end{equation}
We perform inference on $x_{nl}$ and $\gamma_{nl}$, which can also be achieved by using VI with a variational distribution $q(x_{nl}, \gamma_{nl})= q(x_{nl})q(\gamma_{nl})$. 

According to VI, 
\begin{eqnarray}
	q(x_{nl})&\propto &m_{x_{nl}\to f_{x_{nl}}}(x_{nl}) m_{f_{x_{nl}}\to x_{nl}}(x_{nl})\nonumber\\
	&=&\N(x_{nl};\hat x_{nl}, v_{x_{nl}})
\end{eqnarray}
with
$m_{f_{x_{nl}}\to x_{nl}}(x_{nl})$ shown in \eqref{eq:addvi}, and
\begin{eqnarray}
	v_{x_{nl}}=\frac{v_{q_{nl}}}{1+\hat\gamma_{nl}v_{q_{nl}}}, ~ 
	\hat x_{nl}=\frac{\hat q_{nl}}{1+\hat\gamma_{nl}v_{q_{nl}}}.
\end{eqnarray}
The message $m_{f_{x_{nl}}\to \gamma_{nl}}(\gamma_{nl})$ can be expressed as 
\begin{eqnarray}
	m_{f_{x_{nl}}\to \gamma_{nl}}(\gamma_{nl})&\propto& \exp\big( \int_{x_{nl}} q(x_{nl}) \log \mathcal{N}(x_{nl};0,\gamma_{nl}^{-1})\big), \nonumber \\
 &\propto& \gamma_{nl}^{\frac{1}{2}}\exp\big(-\frac{\gamma_{nl}}{2}(|\hat x_{nl}|^2+v_{x_{nl}})\big).
\end{eqnarray}
The message $m_{f_{\gamma_{nl}}\to \gamma_{nl}} (\gamma_{nl})$ is the predefined Gamma distribution, i.e., $m_{f_{\gamma_{nl}}\to \gamma_{nl}} (\gamma_{nl}) \propto \gamma_{nl}^{\epsilon -1}\exp(-\eta\gamma_{nl})$. According to VI,
\begin{eqnarray}
	q(\gamma_{nl}) &\propto& m_{f_{\gamma_{nl}}\to \gamma_{nl}} (\gamma_{nl}) m_{f_{x_{nl}}\to \gamma_{nl}} (\gamma_{nl}) \nonumber \\
	&\propto& \gamma_{nl}^{\epsilon-\frac{1}{2}} \exp\big(-\frac{\gamma_{nl}}{2}(|\hat x_{nl}|^2+v_{x_{nl}}+2\eta)\big). 
\end{eqnarray}
Thus
\begin{eqnarray}
	m_{f_{x_{nl}}\to x_{nl}} (x_{nl}) &\propto& \exp\big( \int_{\gamma_{nl}} q(\gamma_{nl}) \log \mathcal{N}(x_{nl};0,\gamma_{nl}^{-1})\big), \nonumber \\
	&\propto& \mathcal{N} (x_{nl}; 0, \hat \gamma_{nl}^{-1}), \label{eq:addvi}
\end{eqnarray} 
where
\begin{eqnarray}
	\hat\gamma_{nl}=\frac{1+2 \epsilon}{2 \eta+v_{x_{nl}}+|\hat x_{nl}|^2}.
\end{eqnarray}
The above computations for all the entries of $\bX$ can be collectively expressed as
\begin{eqnarray}
&&\bXi_X=\bV_{Q_X}./\left(\mathbf{1}+\hat\bGamma\cdot\bV_{Q_X}\right), \label{eq:SBL1}\\
&&\hat\bX={\hat\bQ_X}./\left(\mathbf{1}+\hat\bGamma\cdot\bV_{Q_X}\right), \\
&&\hat\bGamma=(\mathbf{1}+2 \epsilon \mathbf{1})./\left(2 \eta \mathbf{1}+\bXi_X+|\hat\bX|^{.2}\right),\label{eq:SBL2}
\end{eqnarray}
where $\hat\bGamma$ is a matrix with $\{\gamma_{nl}\}$ as its entries.

We see that UAMP-MF in Algorithm \ref{alg1} involves matrix multiplications and two EVDs per iteration, which dominate its complexity. The complexity of the algorithm is therefore in a cubic order, which is low in a MF problem (the complexity of two matrices product is cubic). In the next section, we will compare the proposed algorithm with state-of-the-art methods in various scenarios. As the algorithms have different computational complexity per iteration and the numbers of iterations required for convergence are also different, we will compare their complexity in terms of runtime, as in the literature. Extensive examples show that UAMP-MF is much faster than state-of-the-art algorithms in many scenarios, and often delivers significantly better performance.



\section{Applications and Numerical Results}

We focus on the applications of UAMP-MF to NFM, RPCA, DL, CSMU and sparse MF. Extensive comparisons with state-of-the-art algorithms are provided to demonstrate its merits in terms of robustness, recovery accuracy and complexity.   

\subsection{UAMP-MF for Robust Principal Components Analysis}
In RPCA \cite{Cand2011}, we aim to estimate a low-rank matrix observed in the presence of noise and outliers. The signal model for RPCA can be written as
\begin{eqnarray}
\bY=\bA\bB+\bE+\bW, \label{eq:RPCA1}
\end{eqnarray}
where the product of a tall matrix $\bA \in \mR^{M\times N}$ and wide matrix $\bB \in \mR^{N\times L}$ is a low-rank matrix, $\bE$ is a sparse outlier matrix, and $\bW$ is noise matrix, which is assumed to be Gaussian and white. To enable the use of UAMP-MF, the model \eqref{eq:RPCA1} can be rewritten as \cite{Parker2014I}
\begin{eqnarray}
{\bY}=
\underbrace{\Big[\bA, \bI]}_{{\bH}}
\underbrace{
\begin{bmatrix}
        \bB \\
        \bE
      \end{bmatrix}}_{\bX}
      +{\bW},
      \label{eq:RPCA2}
\end{eqnarray}
which is a MF problem.

In \eqref{eq:RPCA2}, part of matrix ${\bH}$ is known, and part of matrix ${\bX}$ is sparse. These can be captured by imposing proper priors to the matrix entries.
Similar to \cite{Parker2014I}, for matrix ${\bH}$, the priors are
\begin{eqnarray}
p({h}_{m,n})&=&
\begin{cases}
\N({h}_{m,n}; 0, 1) & n\leq N \\
\N({h}_{m,n}; I_{m,n}, 0) & n\geq N
\end{cases}
\end{eqnarray}
where the entries corresponding to matrix ${\bI}$ have a zero variance, indicating that they are deterministic and known. For the sparse matrix ${\bE}$, the sparsity rate is unknown, and we use the hierarchical Gaussian-Gamma priors. Then we have  
\begin{eqnarray}
p({x}_{n,l})&=&
\begin{cases}
\N({x}_{n,l}; 0, \alpha_x) & n\leq N \\
\N({x}_{n,l}; 0, \gamma^{-1}_{n,l}) & n\geq N
\end{cases}
\end{eqnarray}
where the precisions have Gamma distribution $p(\gamma_{nl})=Ga(\gamma_{nl}; \epsilon, \eta)$. If the knowledge about the parameter $\alpha_x$ is unavailable, it can also be learned using VI, where 
the estimate of $\alpha_x$ can be updated as 
$\hat \alpha_x=||\bXi'_X+|\hat \bX'|^{.2}||^2/NM$ with
\begin{eqnarray}
\bXi'_X&=&{\hat\alpha'_x \bV'_{Q_X}}./{(\hat\alpha'_x \mathbf{1} +\bV'_{Q_X})}, \\
\hat \bX'&=&{\hat\alpha'_x \hat\bQ'_X}./{(\hat \alpha'_x \mathbf{1} +\bV'_{Q_X})},
\end{eqnarray}
where $\bV'_{Q_X}$ and $\hat\bQ'_X$ represent the upper ($n\leq N$) part of matrices $\bV_{Q_X}$ and $\hat\bQ_X$, and $\hat\alpha'_x$ is the estimate of $\alpha_x$ in the last round of iteration. 
In addition, the rank $N$ of matrix $\bZ=\bA\bB$ is normally unknown, and
rank-selection can be performed using the method in \cite{Parker2014II}. 

We compare UAMP-MF with state-of-the-art algorithms including Bi-GAMP \cite{Parker2014I},  IALM \cite{Lin2010} and LMaFit \cite{Wen2012}. The performance is evaluated in terms of the normalized mean squared error (NMSE) of matrix $\bZ$, which is defined as
\begin{eqnarray}
\text{NMSE(Z)} =\frac{\| \bZ- \hat\bH\hat\bX \|^2}{\| \bZ \|^2}.
\end{eqnarray}
In the numerical examples, the low rank matrix $\bZ=\bH\bX$ is generated from $\bH$ and $\bX$ with i.i.d standard Gaussian entries. The sparsity rate of matrix $\bE$ is denoted by $\delta$, and the non-zero entries are independently drawn from a uniform distribution on $[-10,10]$, which are located uniformly at random in $\bE$. It is noted that, with these settings, the magnitudes of the outliers and the entries of $\bZ$ are on the same order, making the detection of the outliers challenging. 
The NMSE performance of various algorithms versus SNR is shown in Fig. \ref{fig:RPCA_MSEvsSNR}, where $\delta=0.1$, $M=L=200$, and the rank $N=50, 70$ and $90$ for (a), (b) and (c), respectively. It can be seen that IALM does not work well, and LMaFit struggles, delivering poor performance when $N=70$ and $90$. Bi-GAMP and UAMP-MF deliver similar performance when $N=50$, but when $N=70$ and $90$, UAMP-MF still works very well while the performance of Bi-GAMP degrades significantly. We also show the performance of UAMP-MF with known rank $N$, which is denoted by UAMP-MF(N) in Fig. \ref{fig:RPCA_MSEvsSNR}, where we see that UAMP-MF has almost the same performance as UAMP-MF(N). Then, with $M=L=200$, we vary the rank $N$ from 30 to 120 and the sparsity rate $\delta$ is changed from 0.05 to 0.3.  
The NMSE of the algorithms with different combinations of $\delta$ and $N$ is shown in Fig.\ref{fig:RPCA}, where SNR = 60 dB.
We see that the blue regions of UAMP-MF are significantly larger than other algorithms, indicating the superior performance of UAMP-MF.

\begin{figure}[!t]
	\centering
	\includegraphics[width=0.5\textwidth]{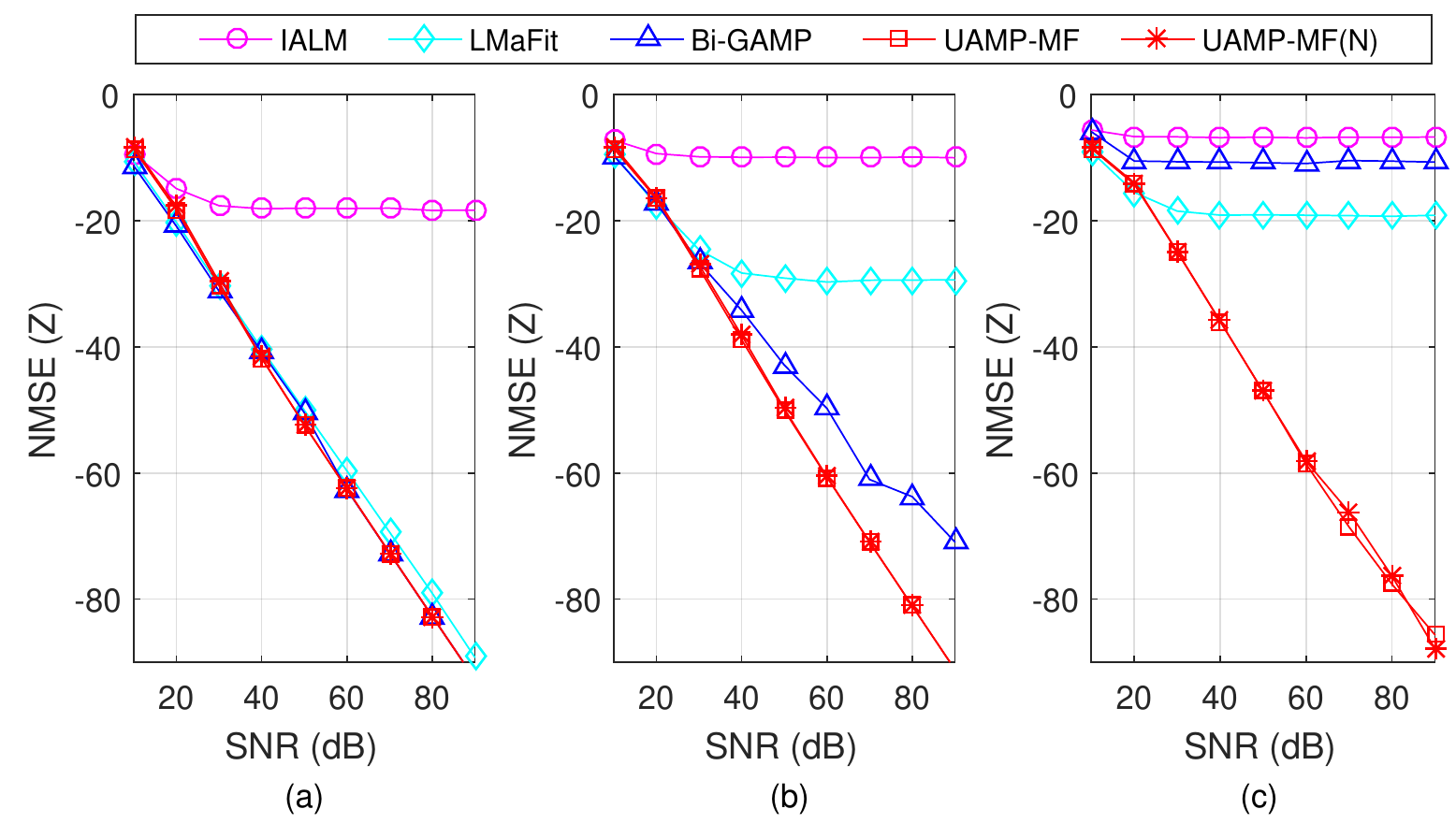}
	\centering
	\caption{NMSE of the algorithms for RPCA, where $\delta=0.1$, and $N=50, 70$ and $90$ in (a), (b) and (c), respectively.
	}
	\label{fig:RPCA_MSEvsSNR}
\end{figure}

\begin{figure}[!t]
	\centering
	\includegraphics[width=0.45\textwidth]{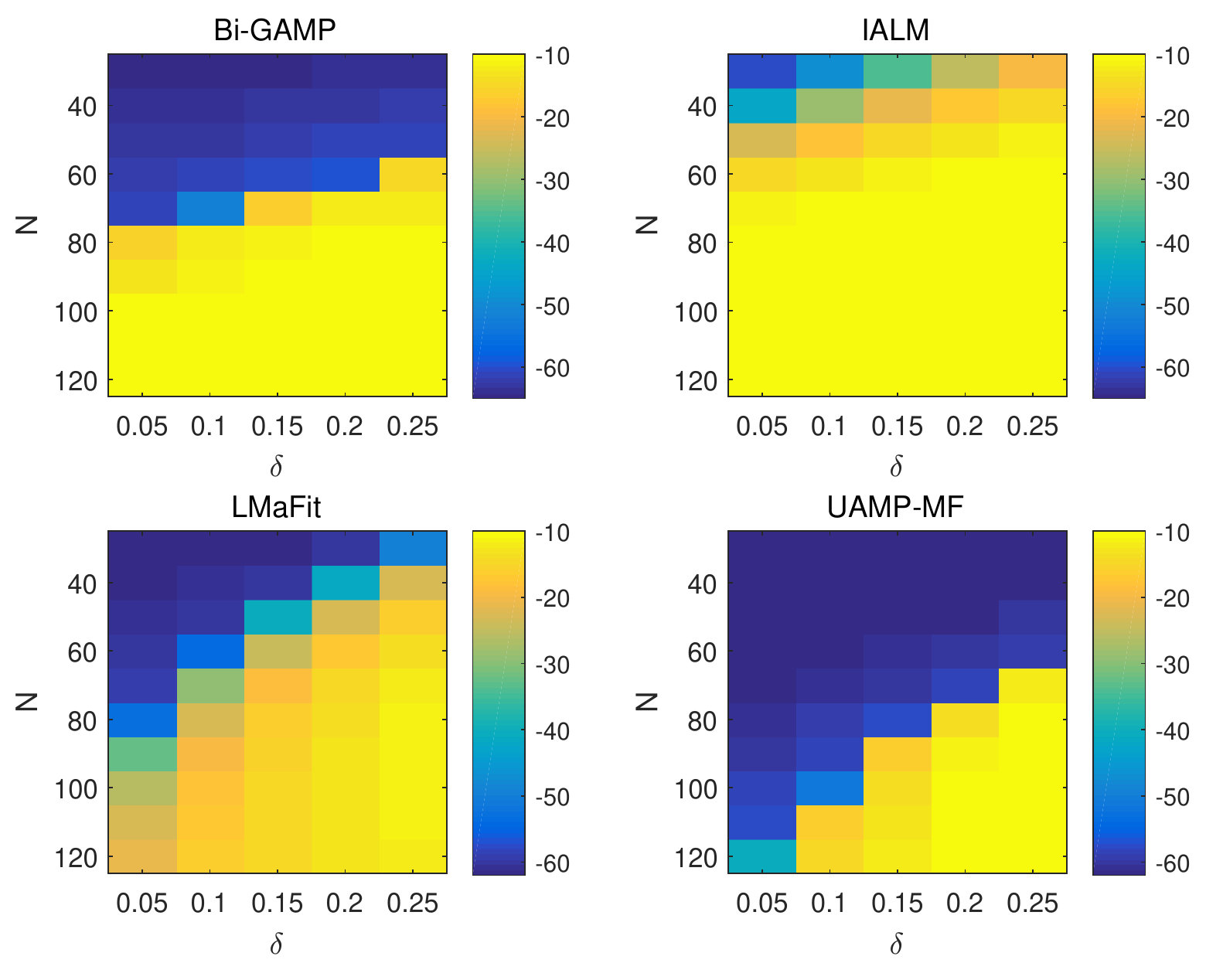}
	\centering
	\caption{NMSE of the algorithms versus the combinations of rank $N$ and sparsity ratio $\delta$ in RPCA.
	}
	\label{fig:RPCA}
\end{figure}

\begin{figure}[!t]
	\centering
	\includegraphics[width=0.4 \textwidth]{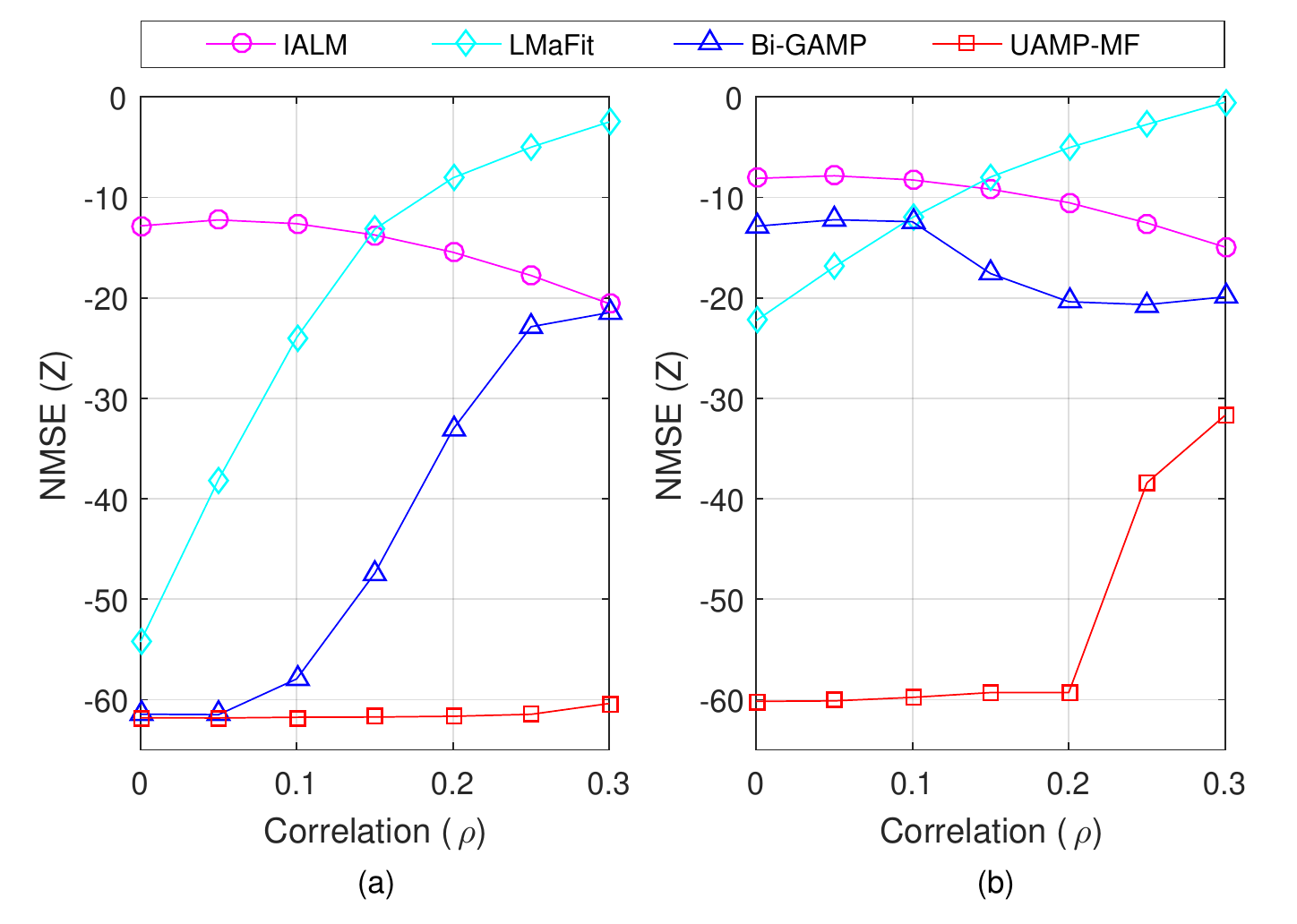}
	\centering
	\caption{{NMSE of the algorithms versus correlation parameter $\rho$ in RPCA, where (a) $N=60$ and (b) $N=80$.}}
	\label{fig:RPCA_MSEvsDelta}
\end{figure}

\begin{figure}[!t]
	\centering
	\includegraphics[width=0.4 \textwidth]{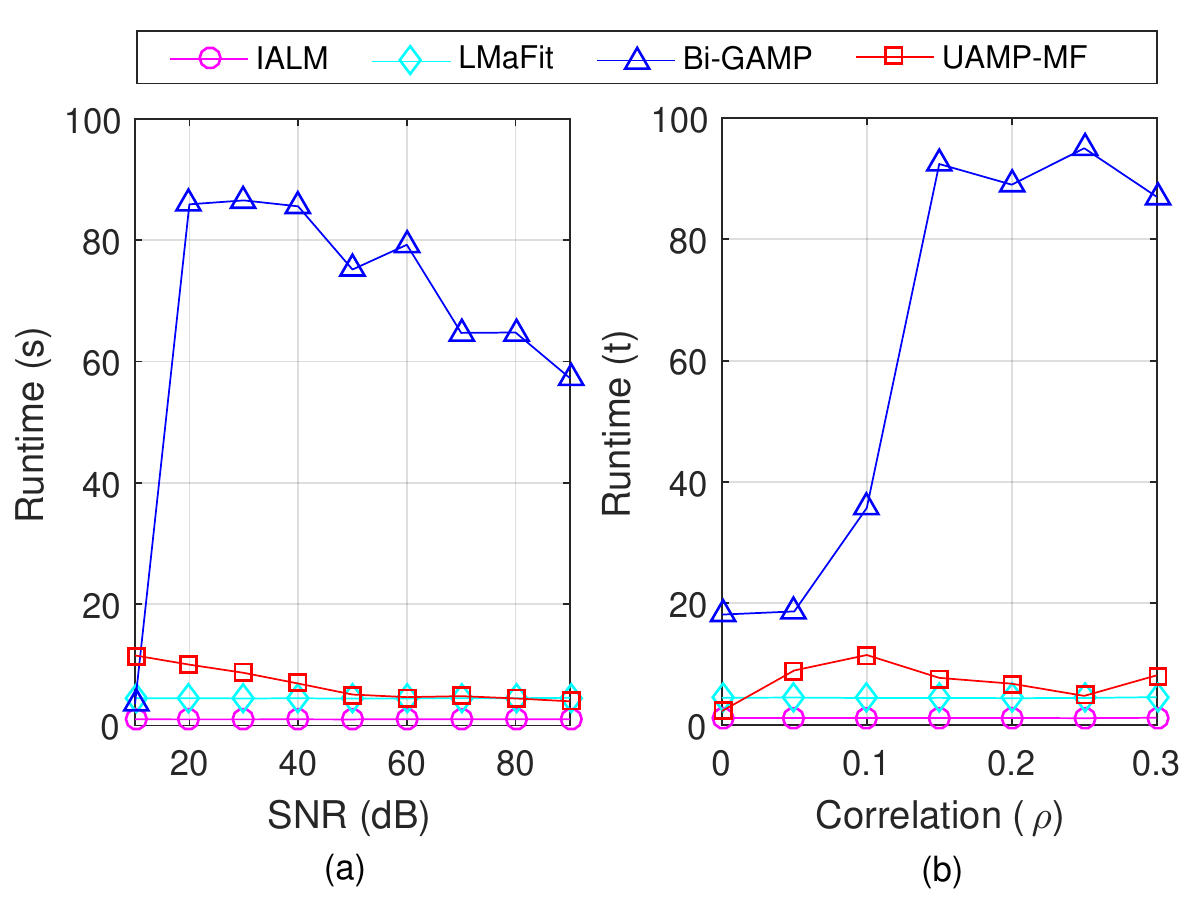}
	\centering
	\caption{Runtime of the algorithms versus (a) SNR and (b) $\rho$.}
	\label{fig:RPCA_Time}
\end{figure}

Next we consider a more challenging case. All the settings are the same as the previous example except that the generated matrices $\bH$ and $\bX$ are no longer i.i.d., but correlated. Take $\bH$ as an example, which is generated with
\begin{eqnarray}
	\bH=\bC_L\bG\bC_R, \label{eq:Correlate}
\end{eqnarray}
where $\bG$ is an i.i.d Gaussian matrix, $\bC_L$ is an $M\times M$ matrix with the $(m,n)$th element given by $\rho^{|m-n|}$ where $\rho\in[0,1]$, and $\bC_R$ is generated in the same way but with size of $N\times N$. The parameter $\rho$ controls the correlation of matrix $\bH$. Matrix $\bX$ is generated in the same way. We show the performance of the algorithms versus the correlation parameter $\rho$ in Fig. \ref{fig:RPCA_MSEvsDelta}, where SNR = 60dB, the rank $N=60$ and $80$ in (a) and (b), respectively. The results show that UAMP-MF is much more robust and it outperforms other algorithms significantly when $\rho$ is relatively large.

The runtime of the algorithms is shown in Fig. \ref{fig:RPCA_Time}, where the simulation parameter settings in Figs. \ref{fig:RPCA_Time} (a) and (b) are the same as those for Fig. \ref{fig:RPCA_MSEvsSNR}(b) and Fig. \ref{fig:RPCA_MSEvsDelta}(a), respectively. The results show that UAMP-MF can be much faster than Bi-GAMP while delivering better performance. IALM and LMaFit is faster than UAMP-MF, but they may suffer from severe performance degradation as shown in Figs. \ref{fig:RPCA_MSEvsSNR} - \ref{fig:RPCA_MSEvsDelta}.   



\subsection{UAMP-MF for Dictionary Learning}
In DL, we aim to find a dictionary $\bH\in\mR^{M\times N}$ that allows the training samples $\bY\in\mR^{M\times L}$ to be coded as
\begin{eqnarray}
\bY=\bH\bX+\bW \label{eq:DL}
\end{eqnarray}
for some sparse matrix $\bX$ and some perturbation $\bW$.
To avoid over-fitting, a sufficiently large number of training examples are needed. For the entries of $\bH$, we use the Gaussian prior as shown in \eqref{adda}, and for the entries of $\bX$, we use the hierarchical Gaussian-Gamma prior (as we assume unknown sparsity rate) as shown in \eqref{adda1} and \eqref{add1a}. 


\begin{figure}[!t]
	\centering
	\includegraphics[width=0.5\textwidth]{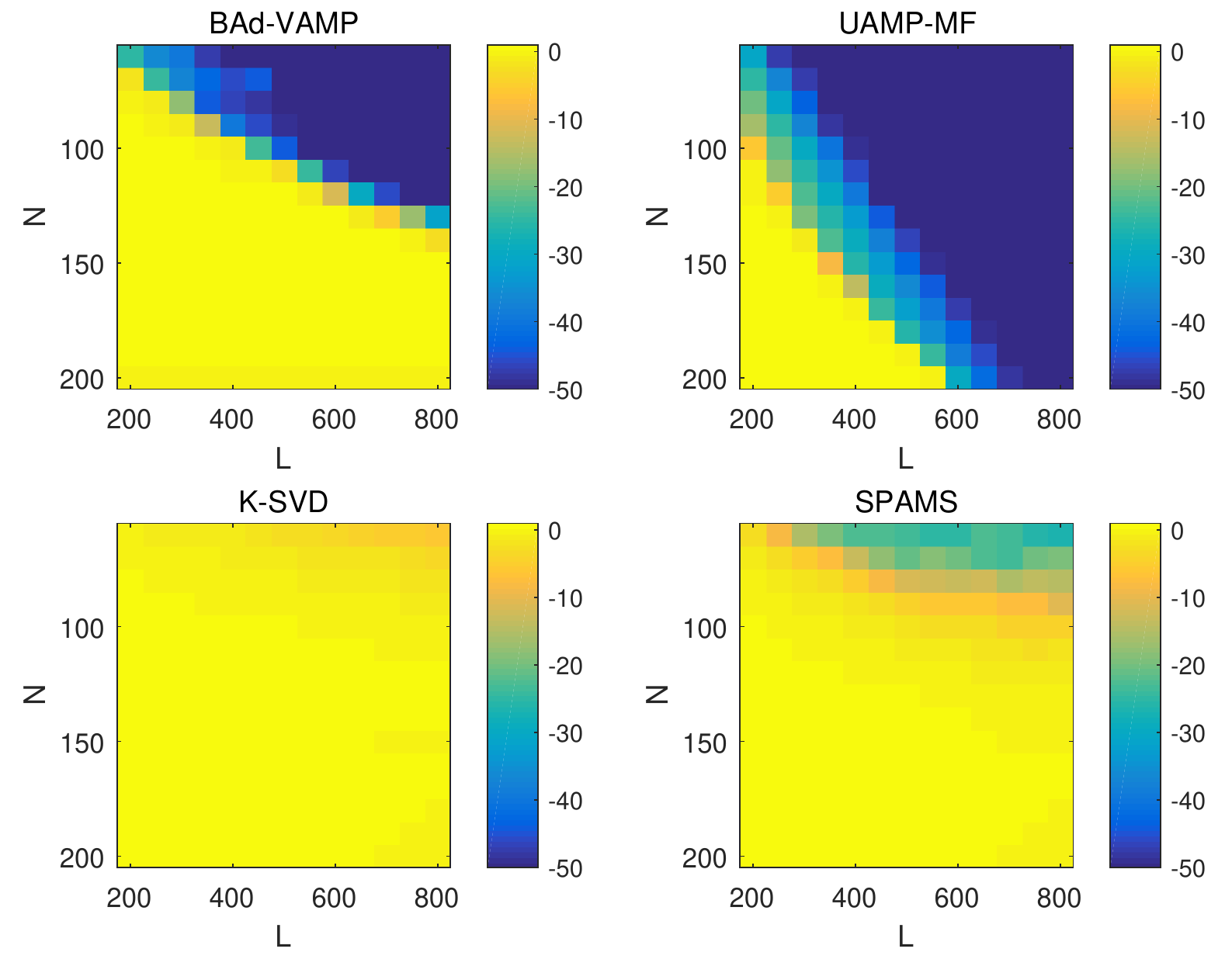}
	\centering
	\caption{NMSE of the algorithms versus combinations of $N$ and $L$ for DL, where $M=N$ and $\rho=0.2$.}
	\label{fig:DL_corr02}
\end{figure}

\begin{figure}[!t]
	\centering
	\includegraphics[width=0.40\textwidth]{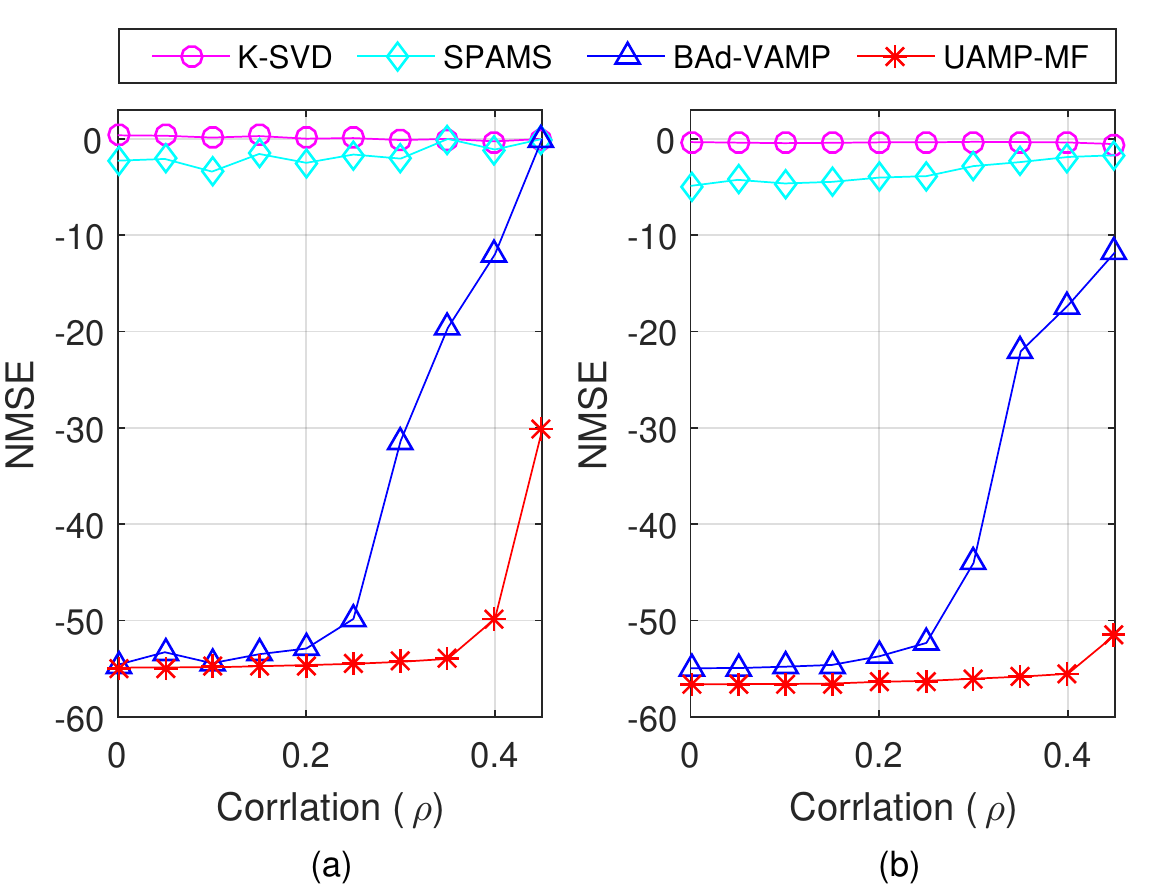}
	\centering
	\caption{NMSE of the algorithms versus $\rho$ for DL, where $M=N=100$, (a) $L=600$  and (b) $L=800$.}
	\label{fig:DL_MSEvsRho}
\end{figure}

We compare UAMP-MF with the state-of-the-art algorithms including BAd-VAMP \cite{Sarkar2019}, K-SVD \cite{Aharon2006} and SPAMS \cite{Mairal2010}. It is worth mentioning that, in the case of DL, compared to Bi-GAMP, BAd-VAMP is significantly enhanced in terms of robustness and performance  \cite{Sarkar2019}. To test the robustness of the algorithms, we generate correlated matrix $\bH$ using \eqref{eq:Correlate} in the numerical examples. 
The sparse matrix  $\bX$ is generated by selecting non-zero entries in each column uniformly and drawing their values from the standard Gaussian distribution, while all other entries are set to zero. The performance of DL is evaluated using the relative NMSE \cite{Sarkar2019}, defined as
\begin{eqnarray}
	\text{NMSE(H)} = \underset{\bJ}{\text{min}}\frac{\| \hat\bH \bJ- \bH \|^2}{\| \bH \|^2},
\end{eqnarray}
where $\bJ$ is a matrix accounting for the permutation and scale ambiguities.


%

\begin{figure}[!t]
	\centering
	\includegraphics[width=0.40\textwidth]{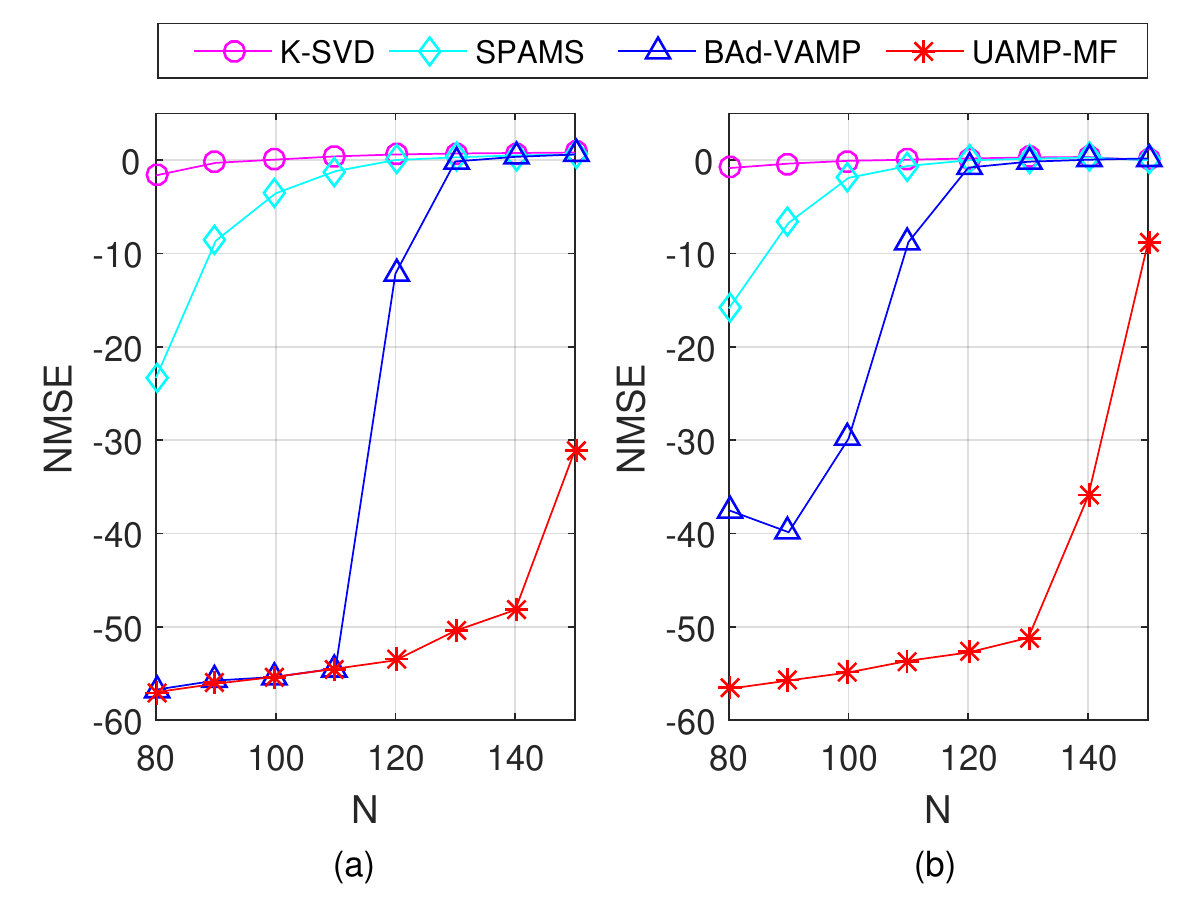}
	\centering
	\caption{{NMSE of the algorithms versus $N$ for DL, where $M=100$, $L=600$, (a) $\rho=0.1$ and (b) $\rho=0.3$.}}
	\label{fig:DL_MSEvsN}
\end{figure}

\begin{figure}[!t]
	\centering
	\includegraphics[width=0.40\textwidth]{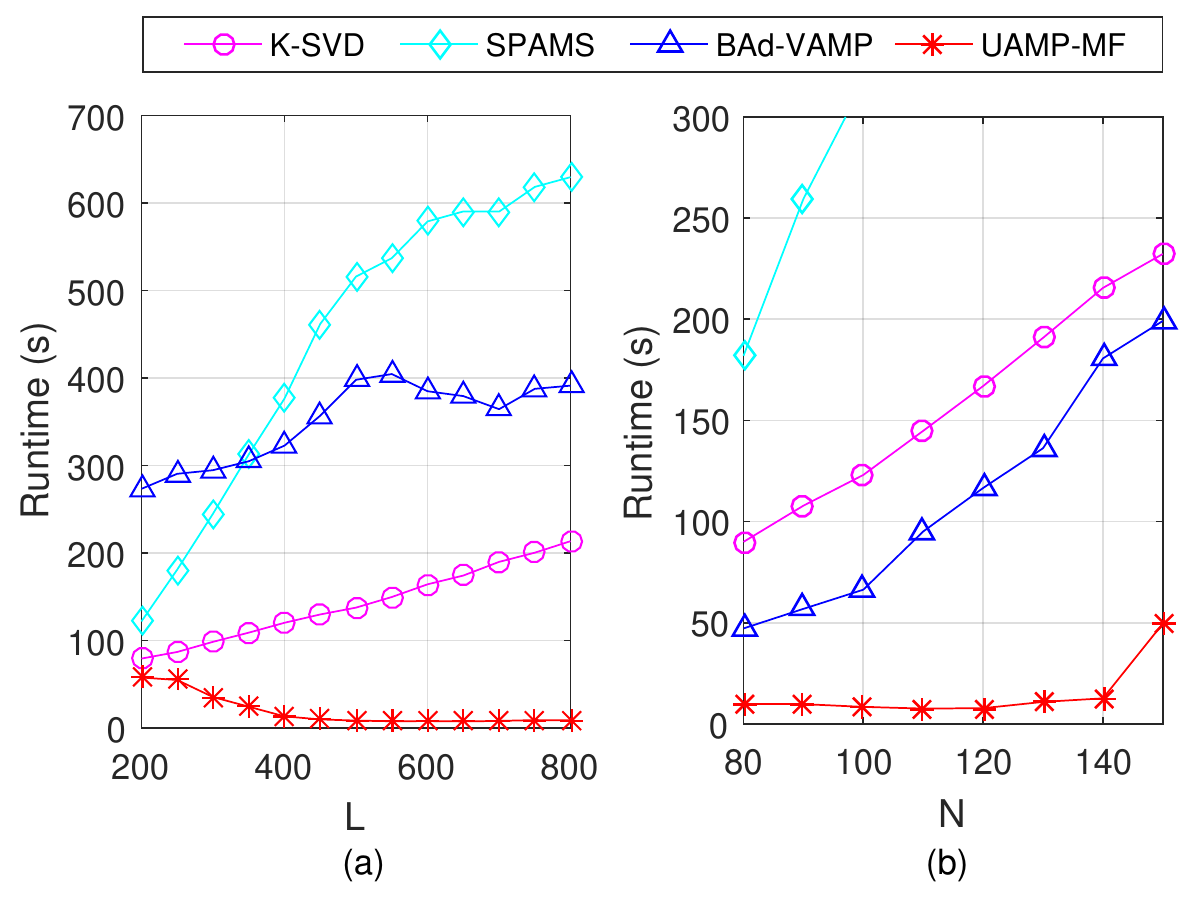}
	\centering
	\caption{Runtime of the algorithms (a) versus $L$, where $M=N=100$ and $\rho=0.1$, and (b) versus $N$, where $M=100$, $L=600$ and $\rho=0.1$ for DL.}
	\label{fig:DL_Time}
\end{figure}


We first assume square dictionaries (i.e., $M=N$) and vary the values of $N$ and $L$. The NMSE performance of the algorithms is shown in Fig.\ref{fig:DL_corr02}, where
SNR = 50dB. It can be seen that, when the correlation parameter $\rho=0.2$, neither K-SVD nor SPAMS has satisfactory performance. In contrast, UAMP-MF and BAd-VAMP works well for a number of combinations of $N$ and $L$, but the blue region of UAMP-MF is significantly larger than that of the BAd-VAMP. 
We show the performance of the algorithms versus the correlation parameter $\rho$ in Fig. \ref{fig:DL_MSEvsRho}, where $N=100$, $L=500$ in (a) and $L=800$ in (b). It can be seen that UAMP-MF exhibits much higher robustness than other algorithms. 

Next, we test the performance of the algorithms when learning an over-complete dictionary, i.e., $N>M$. The setting for the experiments is same as those in the previous one, except that we fix $M$ and $L$, but vary the value of $N$.
The NMSE of the algorithms is shown in Fig. \ref{fig:DL_MSEvsN}, where $M=100$ and $L=600$. The correlation parameter $\rho$ is set to $0.1$ in (a)  and $0.3$ in (b). Again, K-SVD and SPAMS do not work well. Comparing BAd-VAMP with UAMP-MF, we see that UAMP-MF performs significantly better, especially when $\rho=0.3$. 

The runtime versus $L$ of the algorithms is shown in Fig.\ref{fig:DL_Time}.(a), where $M=N=100$ and $\rho=0.1$, and the runtime versus $N$ of the algorithms is shown in Fig. \ref{fig:DL_Time}.(b), where $M=100$, $L=600$, and $\rho=0.1$. The SNR is set to 50 dB. The results show that UAMP-MF can be much faster than other algorithms.


\subsection{UAMP-MF for Compressive Sensing with Matrix Uncertainty}


In practical applications of compressive sensing, we often do not know the measurement matrix $\bH$ perfectly, e.g., due to hardware imperfections and model mismatch. We aim to recover the sparse signal $\bX$ (and estimate the measurement matrix $\bH$) from the noisy measurements $\bY=\bH\bX+\bW$ with
\begin{equation}
	\bH\triangleq \bar \bH+\bH',
\end{equation}
where $\bar \bH$ is a known matrix and $\bH'$ is an unknown disturbance matrix. 
Compared to the case of DL, the difference lies in the prior of $\bH$. We assume that the entries of $\bH'$ are modeled as i.i.d Gaussian distributed variables, i.e., $p(h'_{m,n})=\N(h'_{m,n};0,\nu)$, so the prior of $h_{m,n}$ can be represented as $p(h_{m,n})=\N(h_{m,n}; \bar h_{m,n}, \nu)$. For the entries of $\bX$, we still use the hierarchical Gaussian-Gamma prior. It is worth noting that, if the sparsity of $\bX$ exhibits some pattern, e.g., the columns of $\bX$ share a common support, we can impose that the entries of each row of $\bX$ share a common precision in their Gaussian priors to capture the common support. 

\begin{figure}[!t]
	\centering
	\includegraphics[width=0.45\textwidth]{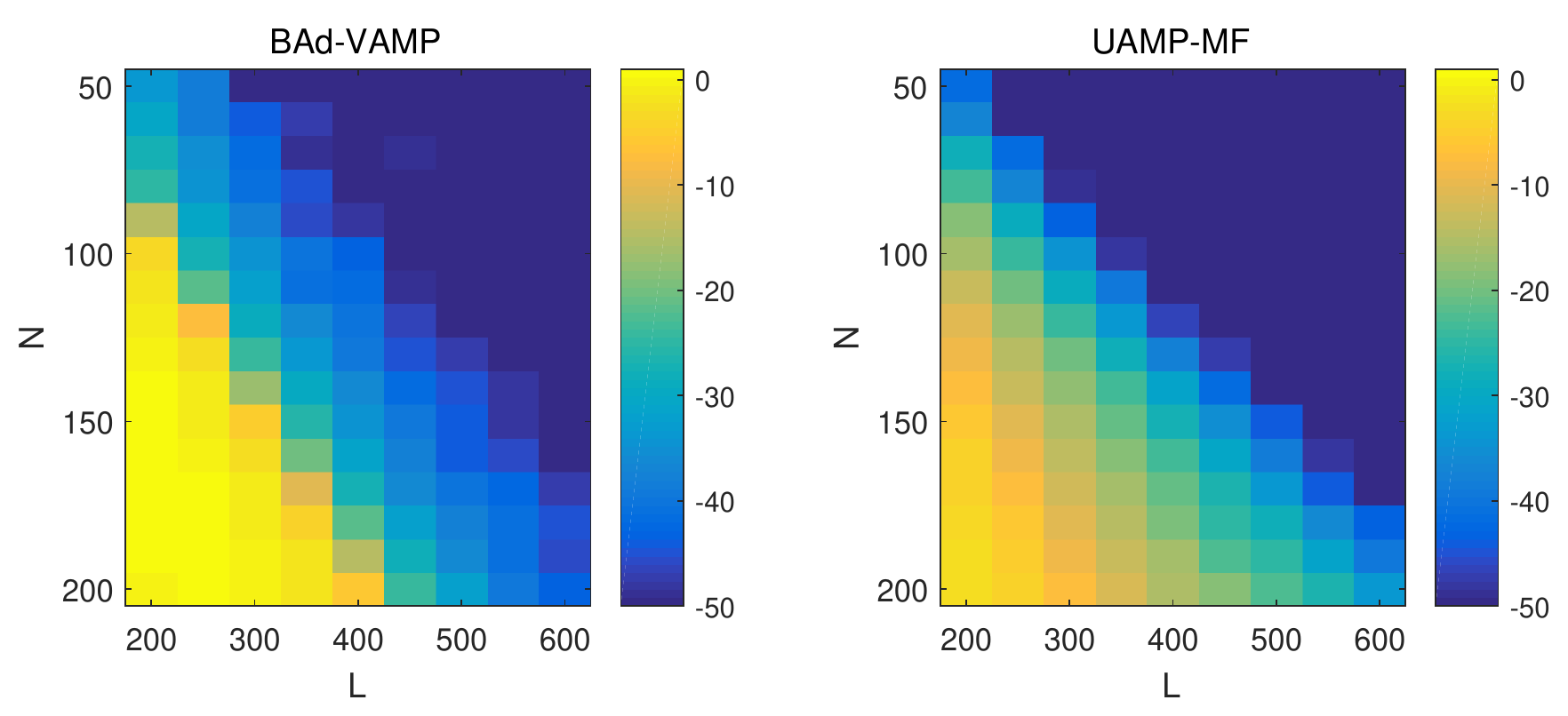}
	\centering
	\caption{NMSE of UAMP-MF and BAd-VAMP for CSMU where SNR = 50dB and $\rho=0.1$.}
	\label{fig:CSMU1}
\end{figure}
\begin{figure}[!t]
	\centering
	\includegraphics[width=0.45\textwidth]{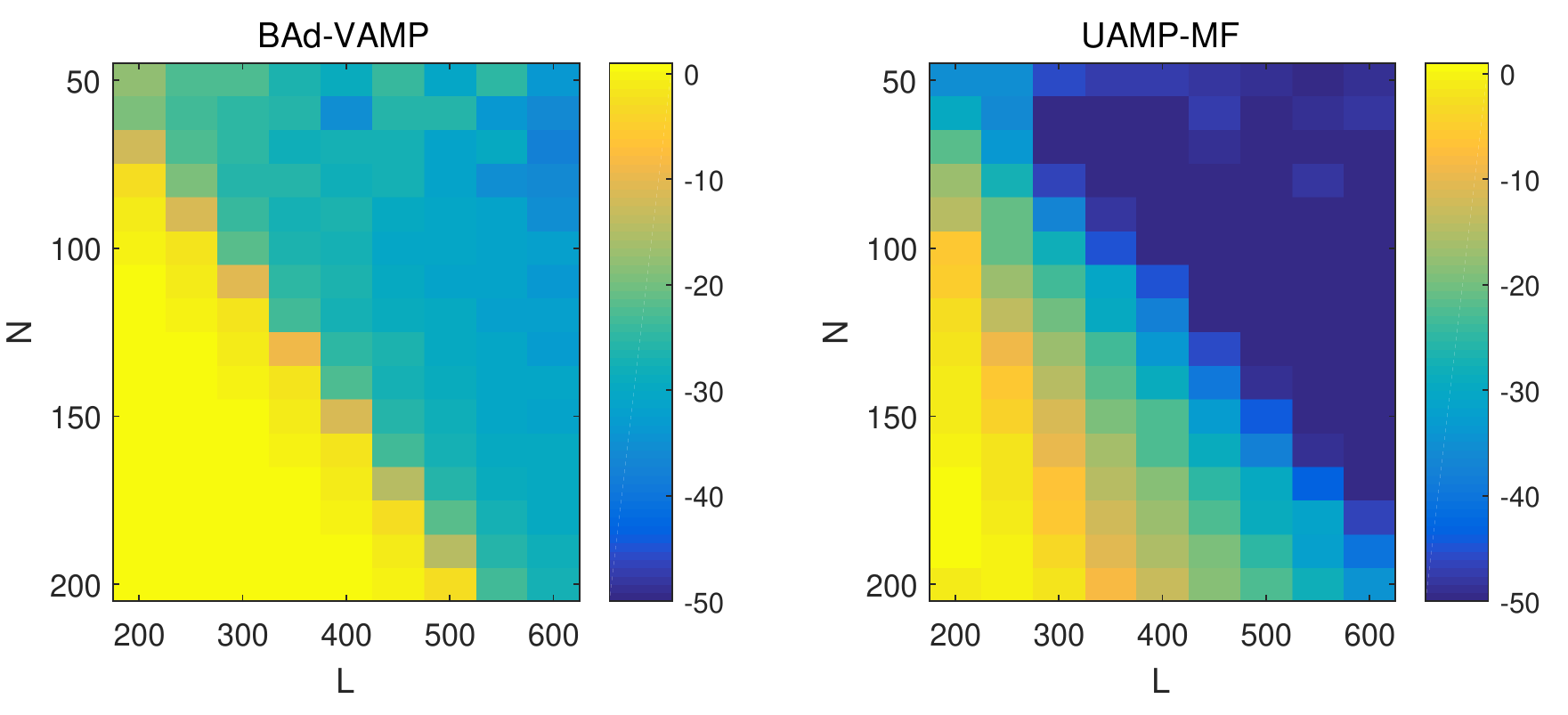}
	\centering
	\caption{NMSE of UAMP-MF and BAd-VAMP for CSMU where SNR = 50dB and $\rho=0.3$.}
	\label{fig:CSMU3}
\end{figure}

\begin{figure}[!t]
	\centering
	\includegraphics[width=0.35\textwidth]{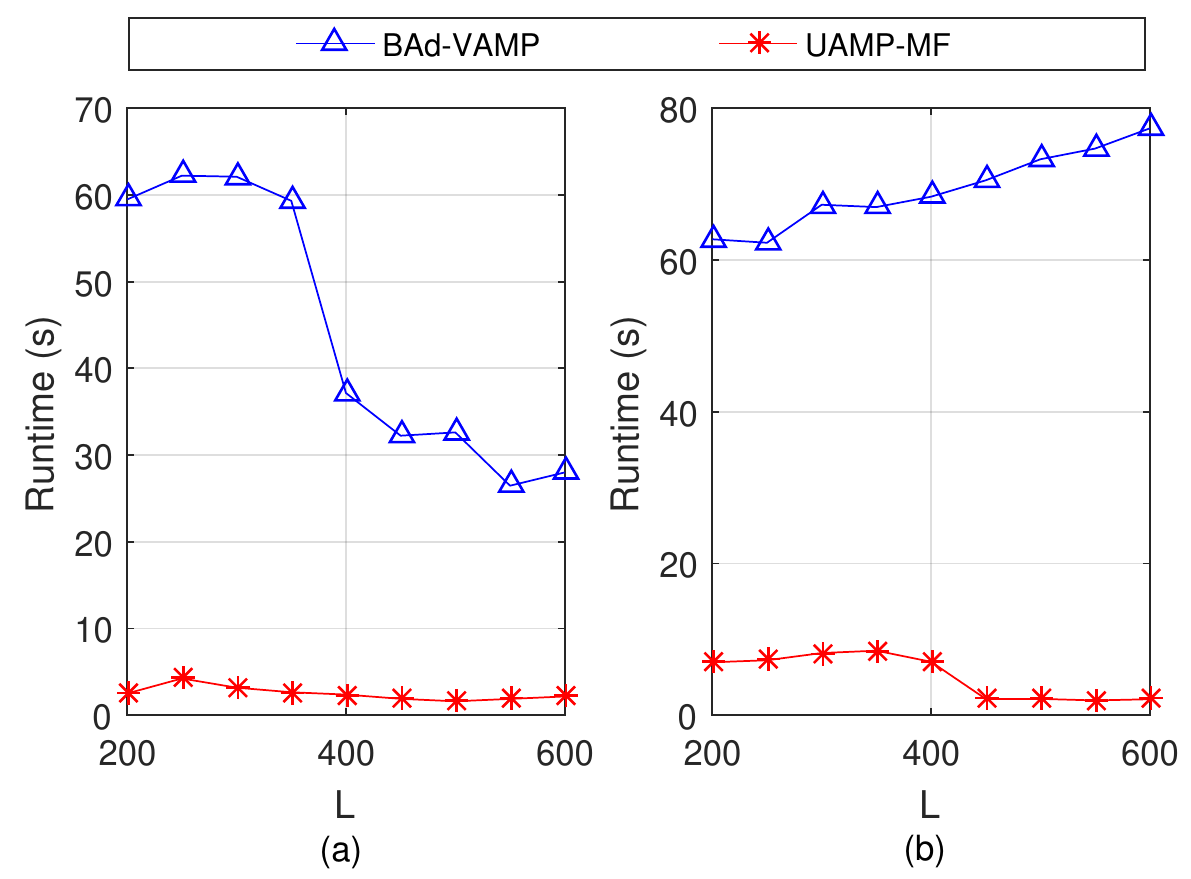}
	\centering
	\caption{Runtime comparison for CSMU (a) $\rho=0.1$, (b) $\rho=0.3$, where $M=N=100$ and SNR = 50dB}
	\label{fig:CSMU1Time}
\end{figure}

In the numerical example, to test the robustness of the algorithms, we generate correlated matrices $\bar \bH$ and {$\bH'$} using \eqref{eq:Correlate}. The sparse matrix $\bX$ is generated in the same way as before. 
It is noted that, different from DL, the ambiguity does not exist thanks to the known matrix $\bH$. The performance is evaluated using the NMSE of $\bX$
\begin{eqnarray}
{\text{NMSE(X)} ={\| \hat\bX- \bX \|^2}/{\| \bX \|^2}.}
\end{eqnarray}

The NMSE performance of UAMP-MF and BAd-VAMP is shown in Fig.\ref{fig:CSMU1} and Fig. \ref{fig:CSMU3} for various combinations of $M=N \in \{50,...,200\}$ and $L \in \{100,...,600\}$. The parameter $\rho=0.1$ and $0.3$ in Figs. \ref{fig:CSMU1} and \ref{fig:CSMU3}, respectively and SNR = 50dB. 
It can be seen that, UAMP-MF outperforms BAd-VAMP significantly when the correlation coefficient $\rho=0.3$. The runtime comparison is shown in Fig. \ref{fig:CSMU1Time}, which indicate that UAMP-MF is much faster.

\subsection{UAMP-MF for Non-Negative Matrix Factorization}
The NMF problem \cite{Lee1999Nature} finds various applications \cite{Dias2012}, \cite{Xu2003}. 
In NMF, from the observation
\begin{eqnarray}
\bY=\bZ+\bW 
\end{eqnarray}
where $\bZ\in \mR_+^{M\times L}$, we aim to find a factorization of $\bZ$ as the product of two non-negative matrices $\bH\in \mR_+^{M\times N}$ and $\bX\in \mR_+^{N\times L}$, i.e., $\bZ\approx \bH \bX $. 
To achieve NMF with UAMP-MF, we set the priors of $p(\bX)$ and $p(\bH)$ to be i.i.d non-negative Gaussian, i.e.,
\begin{eqnarray}
p(x_{n,l})=N_+(x_{n,l};\theta,\phi)
=
\begin{cases}
\frac{\mathcal{N}(x;\theta,\phi)}{\Phi_c(-\theta/\sqrt{\phi})} & x\geq 0 \\
0 & x< 0
\end{cases}\\
p(h_{m,n})=N_+(h_{m,n};\theta,\phi)
=
\begin{cases}
\frac{\mathcal{N}(h;\theta,\phi)}{\Phi_c(-\theta/\sqrt{\phi})} & h\geq 0 \\
0 & h< 0
\end{cases},
\end{eqnarray}
where $\Phi_c$ is the complementary cumulative distribution function.



\begin{figure}[!t]
	\centering
	\includegraphics[width=0.3\textwidth]{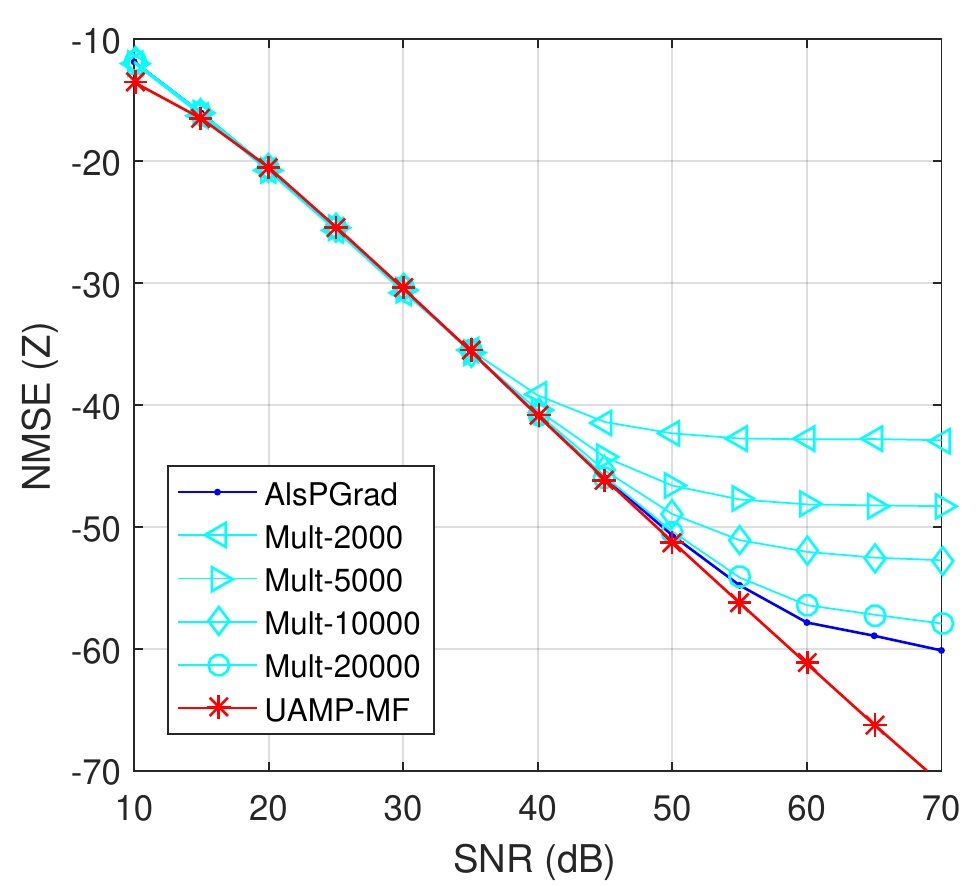}
	\centering
	\caption{{NMSE of the algorithms versus SNR for NMF.}}
	\label{fig:NNMF}
\end{figure}

\begin{figure}[!t]
	\centering
	\includegraphics[width=0.3 \textwidth]{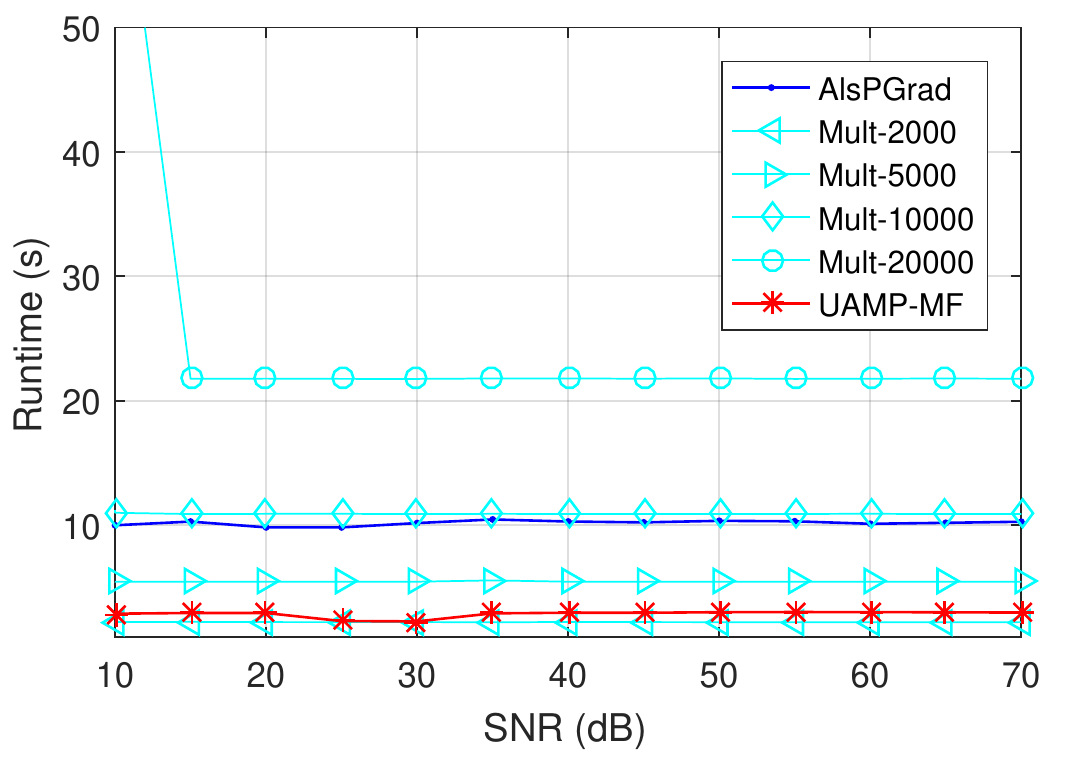}
	\centering
	\caption{{Runtime of the algorithms for NMF.}}
	\label{fig:NNMFtime}
\end{figure}

\begin{figure}[!t]
	\centering
	\includegraphics[width=0.5\textwidth]{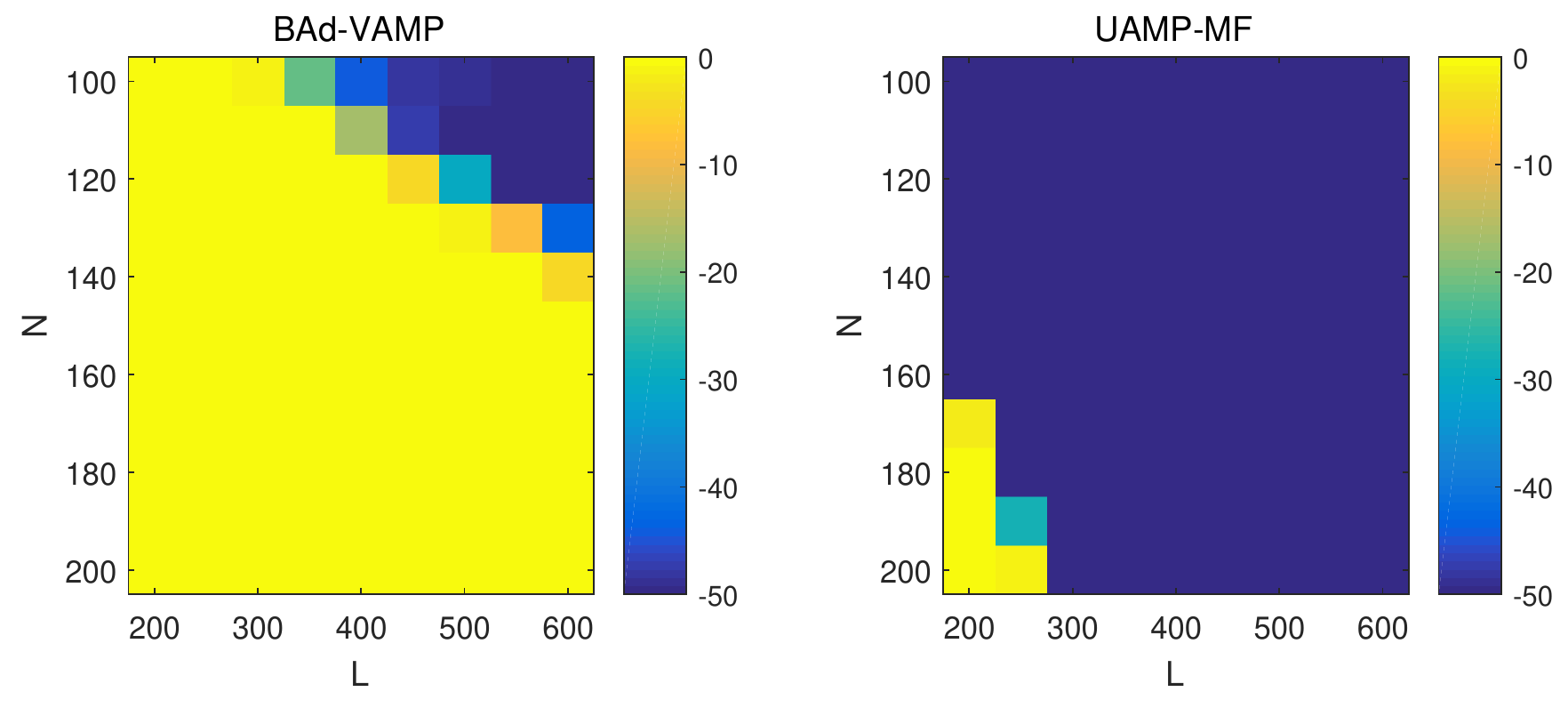}
	\centering
	\caption{
		 NMSE performance comparison for sparse MF.}
	\label{fig:AXsparse}
\end{figure}

We compare UAMP-MF with the alternating non-negative least squares projected gradient algorithm (AlsPGrad) \cite{Lin2007} and multiplicative update algorithm (Mult)\cite{Berry2007}. In the numerical examples, we set $M=L=200$ and $N=100$. 
The metric used to evaluate the performance is defined as
\begin{eqnarray}
\text{NMSE(Z)} = \frac{\| \bZ- \hat\bH\hat\bX \|^2}{\| \bZ \|^2 }.
\end{eqnarray}
For Mult, we use the built-in function `nnmf' in Matlab, and the termination tolerance is set to $10^{-10}$ and maximum number of iterations is set to 2000, 5000, 10000 and 20000 for comparison.

The NMSE results are shown in Fig. \ref{fig:NNMF} and the runtime is shown in  Fig. \ref{fig:NNMFtime}. It can be seen that, with the increase of the maximum number of iterations, the performance of Mult improves considerably, but it comes at the cost of significant increase in runtime. AlsPGrad delivers performance similar to that of Mult with 20000 iterations. UAMP-MF delivers the best performance while with significant smaller runtime compared to Mult (except Mult-2000) and AlsPGrad.

\begin{figure}[!t]
	\centering
	\includegraphics[width=0.3\textwidth]{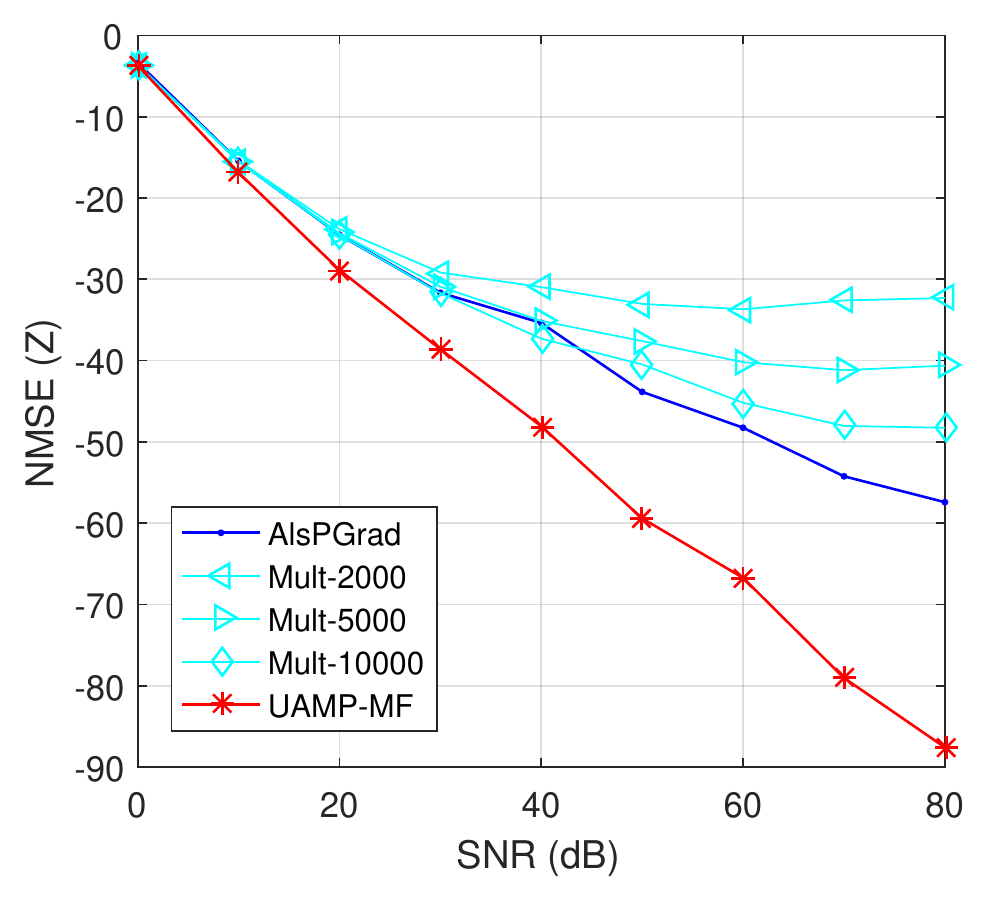}
	\centering
	\caption{Performance comparison for sparse NMF.} 
\label{fig:NMF_Sparse}
\end{figure}

\begin{figure}[!t]
\centering
\includegraphics[width=0.35\textwidth]{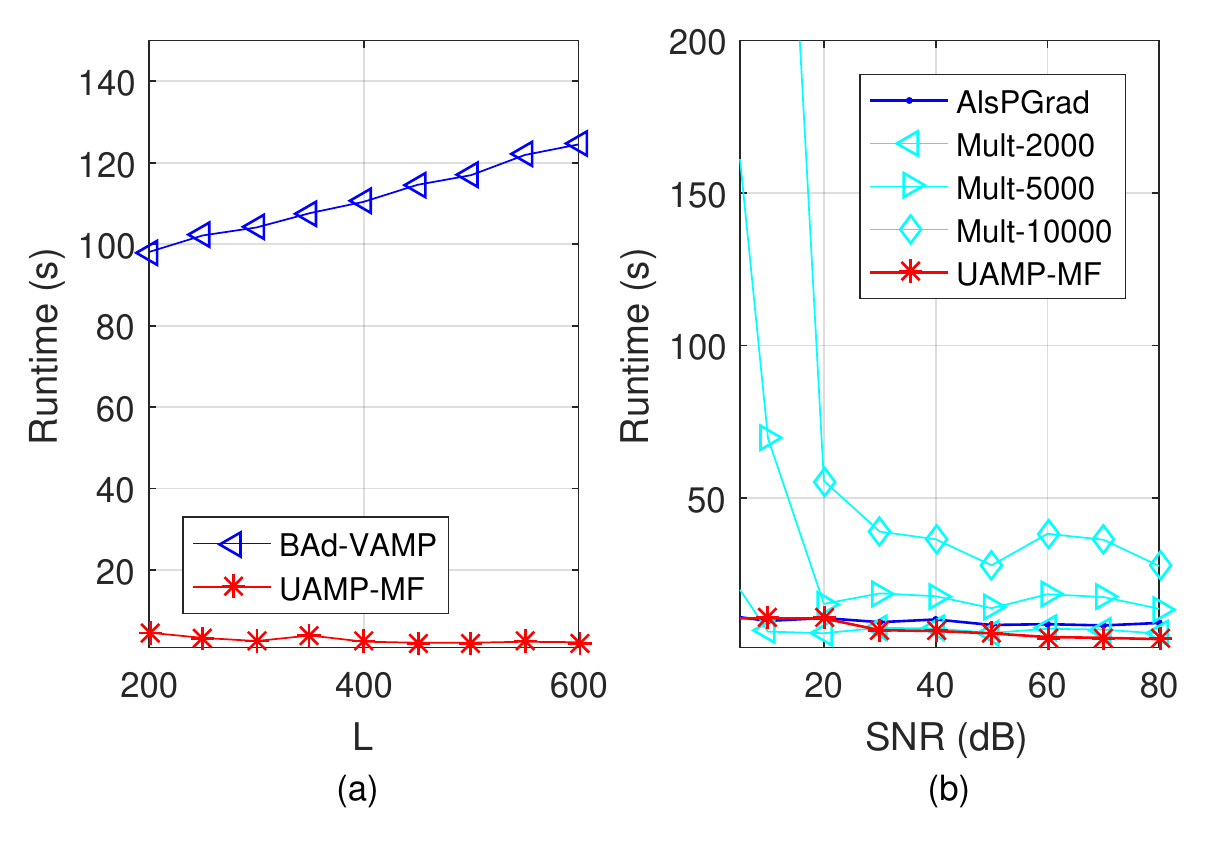}
\centering
\caption{Runtime comparison (a) sparse MF; (b) sparse NMF.}
\label{fig:AXsparseTime}
\end{figure}

\subsection{UAMP-MF for Sparse Matrix Factorization}
We consider the use of UAMP-MF to solve the sparse MF problem \cite{Hoyer2002Non}. From the observation
\begin{eqnarray}
	\bY=\bZ+\bW, 
\end{eqnarray}
where $\bZ\in \mR^{M\times L}$, and we aim to find a factorization of $\bZ$ as the product of two sparse matrices $\bH\in \mR^{M\times N}$ and $\bX\in \mR^{N\times L}$, i.e., $\bZ\approx \bH \bX $. 


To find the sparse matrices $\bH$ and $\bX$ with UAMP-MF, we use the sparsity promoting hierarchical Gaussian-Gamma prior for entries in both matrices, 
i.e.,
\begin{eqnarray}
p(\bH)\!\!\!&=&\!\!\!\prod_{m,n}p(h_{m,n})=\prod_{m,n}N(h_{m,n};0,\gamma_{h_{m,n}}^{-1}) \\
p(\bX)\!\!\!&=&\!\!\!\prod_{n,l}p(x_{n,l})=\prod_{n,l}N(x_{n,l};0,\gamma_{x_{n,l}}^{-1}) \\
p(\gamma_{h_{m,n}})\!\!\!&=&\!\!\!Ga(\gamma_{h_{m,n}};\epsilon,\eta) \\
p(\gamma_{x_{n,l}})\!\!\!&=&\!\!\!Ga(\gamma_{x_{n,l}};\epsilon,\eta).
\end{eqnarray}
The computations of the a posterior means and variances of the entries of $\bH$ and $\bX$ are the same as \eqref{eq:SBL1}-\eqref{eq:SBL2}.
We compare the performance of UAMP-MF and BAd-VAMP, and the results are shown in Fig.\ref{fig:AXsparse} for various combinations of $N\in \{50,...,200\}$ and $L\in \{100,...,800\}$, where $M=N$, SNR = 50dB, and the sparsity rate $\delta$ of $\bH$ and $\bX$ is set to $0.2$. It can be seen that the performance of UAMP-MF is significantly better than that of BAd-VAMP as no priors can be imposed on $\bH$ in BAd-VAMP.  

We can also achieve sparse NMF with UAMP-MF, where $\bZ\in \mR_+^{M\times L}$ and a sparse non-negative factorization $\bZ\approx \bH \bX $ is to be found. In this case, we can employ a non-negative Bernoulli-Gaussian distribution for the entries in both $\bH$ and $\bX$. We compare UMAP-MF with Mult and AlsPGrad and the numerical results are shown in Fig. \ref{fig:NMF_Sparse}, where the sparsity rate $\delta=0.1$ for both $\bH$ and $\bX$, $M=200$, $N=100$ and $L=200$. We can see that the performance of UAMP-MF is much better than that of other algorithms as they lack a sparsity promoting mechanism. 

The runtime comparisons for sparse MF and spares NMF are shown in Fig. \ref{fig:AXsparseTime}. It can be seen that UAMP-MF is much faster than BAd-VAMP. Compared to AlsGrad and Mul with 2000 iterations, UAMP-MF has similar runtime but delivers much better performance. 




\section{Conclusions}

In this paper, leveraging variational inference and UAMP, we designed a message passing algorithm UAMP-MF to deal with matrix factorization, where UAMP is incorporated into variational inference through a whitening process. By imposing proper priors, UAMP-MF can be used to solve various problems, such as RPCA, DL, CSMU, NMF, sparse MF and sparse NMF. Extensive numerical results demonstrate the superiority of
UAMP-MF in recovery accuracy, robustness, and computational complexity, compared to state-of-the-art algorithms.

\begin{appendices}

\section{Proof of Proposition 1}

	According to VI, the message from $f_Y$ to $\bX$ can be expressed as 	
	\begin{eqnarray}
		&&\!\!\!\!\!\!\!\!\!\!m_{f_Y\to X}(\bX) \nonumber\\
		&&\!\!\!\!\!\!\!\!\!\!\propto \exp\left(\int_{\bH, \lambda} q(\bH)q(\lambda) \log f_Y   \right) \nonumber\\
		&&\!\!\!\!\!\!\!\!\!\!\propto \exp\Big(-\hat\lambda\int_{\bH} \Tr\Big( (\bY-\bH\bX)\tra (\bY-\bH\bX)\Big) q(\bH) \Big)\nonumber\\
		&&\!\!\!\!\!\!\!\!\!\!\propto \exp\left(-\hat\lambda \int_{\bH}(\by-\tVec(\bH\bX))\tra (\by-\tVec(\bH\bX)) q(\bH) \right)\nonumber\\
		&&\!\!\!\!\!\!\!\!\!\!= \exp\left(-\hat\lambda \int_{\bh}(\by-\tilde\bX\bh))\tra (\by-\tilde\bX\bh)) q(\bh) \right)\label{eq:msg_fyX_proof1}\\
		&&\!\!\!\!\!\!\!\!\!\!=\exp\Big(-\hat\lambda \int_{\bh}\big(\by\tra \by+\underbrace{\bh\tra\tilde\bX\tra\tilde\bX\bh}_{(i)}  \nonumber \\
		&&\ \ \ \ \ \ \ \ \ \ \ \ \ \ \ \ \ \ \ \ \ \ \ \ \ \ \   -\underbrace{\bh\tra\tilde\bX\tra\by}_{(ii)} -\underbrace{\by\tra\tilde\bX\bh}_{(iii)}\big) q(\bh)\Big), \label{eq:msg_fyX_proof}
	\end{eqnarray}
	where
	$\bh=\text{Vec}(\bH)$, $\by=\text{Vec}(\bY)$ and $\tilde\bX=\bX\tra\otimes \bI_M$. In the derivation of \eqref{eq:msg_fyX_proof1}, we use the matrix identity $\text{Vec}(\bH\bX) =\text{Vec}(\bI_M\bH\bX)= \left(\bX\tra\otimes \bI_M\right) \bh$ \cite{matrixcook}.
	
	Next, we work out the the integration of the three terms $(i)$, $(ii)$ and $(iii)$ in \eqref{eq:msg_fyX_proof}. For term $(i)$, we have 
	\begin{eqnarray}
		&&\!\!\!\!\!\!\!\!\!\int_{\bh}\bh\tra\tilde\bX\tra\tilde\bX\bh q(\bh)\nonumber\\
		&&\!\!\!\!\!\!\!\!\!=\int_{\bh}\Tr\left(\tilde\bX\tra\tilde\bX\bh\bh\tra \right) q(\bh)\nonumber\\
		&&\!\!\!\!\!\!\!\!\!=\Tr\left(\tilde\bX\tra\tilde\bX(\hat\bh\hat\bh\tra+\bV_H\otimes\bU_H)\right)\nonumber\\
		&&\!\!\!\!\!\!\!\!\!=\Tr(\hat\bh\tra\tilde\bX\tra\tilde\bX\hat\bh)
		+\Tr(\tilde\bX(\bV_H\otimes\bU_H)\tilde\bX\tra)\nonumber\\
		&&\!\!\!\!\!\!\!\!\!=\Tr(\tVec(\hat\bH)\tra(\bX\tra\otimes\bI_M)\tra(\bX\tra\otimes\bI_M)\tVec(\hat\bH))\nonumber\\
		&&\ \ \ \ \ \ \ +\Tr((\bX\tra\otimes \bI_M)(\bV_H\otimes\bU_H)(\bX\tra\otimes \bI_M))\nonumber\\
		&&\!\!\!\!\!\!\!\!\!=\Tr\big(\tVec(\hat\bH\bX)\tra \tVec(\hat\bH\bX)\big)\nonumber\\
		&&\ \ \ \ \ \ \ \ +\Tr\big((\bX\otimes \bI_M)(\bX\tra\otimes \bI_M)(\bV_H\otimes\bU_H)\big)\label{eq:msg_fyX_parti_1}\\
		&&\!\!\!\!\!\!\!\!\!=\Tr(\bX\tra\hat\bH\tra\hat\bH\bX)\nonumber\\
		&&\ \ \ \ \ \ \ \ +\Tr(((\bX\bX\tra)\otimes\bI_M)(\bV_H\otimes \bU_H))\label{eq:msg_fyX_parti_2}\\
		&&\!\!\!\!\!\!\!\!\!=\Tr(\bX\tra\hat\bH\tra\hat\bH\bX)+\Tr(\bU_H)\Tr(\bX\bX\tra\bV_H)
		\label{eq:msg_fyX_parti_3}\\
		&&\!\!\!\!\!\!\!\!\!=\Tr(\bX\tra(\hat\bH\tra\hat\bH+\Tr(\bU_H)\bV_H)\bX), \label{eq:msg_fyX_parti}
	\end{eqnarray}
	where we use the matrix identity $\Tr\left(\bA\bB\bC\right)=\Tr\left(\bC\bA\bB\right)$ in deriving \eqref{eq:msg_fyX_parti_1},  $\left(\bA\otimes\bB\right)\left(\bC\otimes\bD\right)=(\bA\bB)\otimes(\bC\bD)$ and $\left(\bA\otimes \bB\right)\tra=\bA\tra\otimes\bB\tra$ in deriving \eqref{eq:msg_fyX_parti_2}, and $\Tr(\bA\otimes\bB)=\Tr(\bA)\Tr(\bB)$ in deriving \eqref{eq:msg_fyX_parti_3}.  

	Regarding term $(ii)$, we have
	\begin{eqnarray}
		&&\int_{\bh}\bh\tra\tilde\bX\tra\by q(\bh)\nonumber\\
		&&=\int_{\bH}\text{Vec}(\bH)\tra(\bX\tra\otimes \bI_M)\tra\text{Vec}(\bY)q(\bH)\nonumber\\
		&&=\int_{\bH}\Tr(\bX\tra\bH\tra\bY)q(\bH) \nonumber\\
		&&=\Tr\big(\bX\tra\hat\bH\tra\bY\big). \label{eq:msg_fyX_partii}
	\end{eqnarray}
	Similarly,  term $(iii)$ can be expressed as
	\begin{eqnarray}
		\int_{\bh}\by\tra\tilde\bX\bh q(\bh)=\Tr(\bY\tra\hat\bH\bX). \label{eq:msg_fyX_partiii}
	\end{eqnarray}
Based on the above results, the message
	\begin{eqnarray}
		&&\!\!\!\!\!\!\!\!\!\!\!\!\!\!\!m_{f_Y\to X}(\bX) \propto \exp\Big(-\hat\lambda \Tr\big(\bX\tra(\hat\bH\tra\hat\bH+\Tr(\bU_H)\bV_H)\bX\nonumber\\
		&&\ \ \ \ \ \ \ \ \ \ \ \ \ \ \ \ \ \ -\bX\tra\hat\bH\tra\bY-\bY\tra\hat\bH\bX+\bY\tra\bY\big) \Big). 
	\end{eqnarray}
	Comparing the result against the matrix normal distribution \eqref{eq:MN_Distrib}, we have the result shown in \eqref{eq:msg_fyX_result}. 

\section{Proof of Proposition 2}

	According to VI, the message $m_{f_Y \to H}(\bH)$ is computed as
	\begin{eqnarray}
		&&m_{f_Y\to H}(\bH) \nonumber \\
		&&\propto \exp\left(\int_{\bX, \lambda} q(\bX) q(\lambda) \log f_Y \right)\nonumber\\
		&&\propto\exp\left(-\hat\lambda\int_{\bX} \Tr( (\bY-\bH\bX)\tra (\bY-\bH\bX)) b(\bX) \right)\nonumber\\
		&&=\exp\left(-\hat\lambda \int_{\bx} (\by-\tilde\bH\bx)\tra (\by-\tilde\bH\bx)b(\bx)\right)\nonumber\\
		&&=\exp\Big(-\hat\lambda \int_{\bx} \big(\by\tra \by+ \bx\tra\tilde\bH\tra\tilde\bH\bx\nonumber\\
		&&\ \ \ \ \ \ \ \ \ \ \ \ \ \ \ \ \ \ \ \ \ \ \  \ \ \ \  -\bx\tra\tilde\bH\tra\by  -\by\tra\tilde\bH\bx \big) b(\bx) \Big), \label{eq:msg_fyH_proof}
	\end{eqnarray}
	where {$\tilde\bH\triangleq \bI_L\otimes \bH$}. Similar to the derivation of \eqref{eq:msg_fyX_parti}, the integration of the terms in \eqref{eq:msg_fyH_proof} can be expressed as
	\begin{eqnarray}
		\int_{\bx}\bx\tra\tilde\bH\tra\by b(\bx)= \Tr\big(\hat\bX\tra\bH\tra\bY\big)
	\end{eqnarray}
	and
	\begin{eqnarray}
		&&\int_{\bx}\bx\tra\tilde\bH\tra\tilde\bH\bx b(\bx) \nonumber\\
		&&= \int_{\bx} \Tr\left(\tilde\bH\tra\tilde\bH\bx\bx\tra\right)b(\bx)\nonumber\\
		&& =\Tr\big(\tilde\bH\tra\tilde\bH(\hat\bx\hat\bx\tra+\bV_X\otimes \bU_X)\big)\nonumber\\
		&&=\Tr\big(\hat\bX\tra\bH\tra\bH\hat\bX\big)+\Tr\big(\bV_X\otimes(\bH\tra\bH\bU_X)\big)\nonumber\\
		&&=\Tr\big(\hat\bX\tra\bH\tra\bH\hat\bX\big)+\Tr(\bV_X)\Tr\big(\bH\bU_X\bH\tra\big)\nonumber\\
		&&=\Tr\big(\bH(\hat\bX\hat\bX\tra+\Tr(\bV_X)\bU_X)\bH\tra\big).
	\end{eqnarray}
The message from $f_Y$ to $\bH$ can be represented as
	\begin{eqnarray}
		&&m_{f_Y\to H}(\bH)\nonumber\\
		&&\propto\exp\Big(-\hat\lambda \Tr(\bY\bY\tra+\bH(\hat\bX\hat\bX\tra+\Tr(\bV_X)\bU_X)\bH\tra\nonumber\\
		&&\ \ \ \ \ \ \ \ -\bH\hat\bX\bY\tra-\bY\hat\bX\tra\bH\tra) \Big).
	\end{eqnarray}
Comparing the above against the matrix normal distribution, we obtain the result shown by \eqref{eq:msg_fyH_result} - \eqref{eq:msg_fyH_V}.

\section{Proof of Proposition 3}
	According to VI, the message
	\begin{eqnarray}
		&&m_{f_Y\to \lambda}(\lambda) \nonumber \\
		&& =\text{det}(\lambda^{-1}\bI_M\otimes\bI_L)\exp\Big(-\lambda \big(\text{Vec}
		(\bY)-\tVec(\bH\bX)\big)\tra \nonumber\\
		&&\ \ \ \ \ \ \ \ \ \ \ \ \ \ \ \ \ \ \ \ \ \ \ \ \ \ \ \ \ \ \ \ \ \ \ \ \   \big(\tVec(\bY)-\tVec(\bH\bX)\big)\Big)\nonumber\\
		&&=\lambda^{ML}\exp\Big(-\lambda\int_{\bH,\bX}\Tr((\bY-\bH\bX)\tra\nonumber\\
		&&\ \ \ \ \ \ \ \ \ \ \ \ \ \ \ \ \ \ \ \ \ \ \ \ \ \ \  \ \ \ \ \ \ \  (\bY-\bH\bX))b(\bH)b(\bX)\Big)\nonumber\\
		&&=\lambda^{ML}\exp\Big(-\lambda C \Big),
	\end{eqnarray}
	where
	\begin{eqnarray}
		&&\!\!\!\!\!\!\!\!\! C=\int_{\bH,\bX}
		\Tr((\bY-\bH\bX)\tra(\bY-\bH\bX))q(\bH)q(\bX)\nonumber\\
		&&\!\!\!\!\!\!\!\!\! \!=\!\int_{\bx,\bH} (\by\tra\by+\bx\tra\tilde\bH\tra\tilde\bH\bx-\bx\tra\tilde\bH\tra\by-\by\tra\tilde\bH\bx)
		q(\bx)q(\bH)\nonumber\\
		&&\!\!\!\!\!\!\!\!\! \!=\!\int_{\bH} \Tr(\bY\bY\tra+\bH(\hat\bX\hat\bX\tra+\Tr(\bV_X)\bU_X)\bH\tra\nonumber\\
		&&\ \ \ \ \ \ \ \ \ \ \ \ \ \ \ \ \ \ \ \ \ \ \ \ \ \ \ \  -\bH\hat\bX\bY\tra-\bY\hat\bX\tra\bH\tra)q(\bH)\nonumber\\
		&&\!\!\!\!\!\!\!\!\!\! =\Tr(\bY\bY\tra -\hat\bH\hat\bX\bY\tra-\bY\hat\bX\tra\hat\bH\tra)\nonumber\\
		&& +\Tr((\hat\bX\hat\bX\tra+\Tr(\bV_X)\bU_X)(\hat\bH\tra\hat\bH+\Tr(\bU_H)\bV_H)\tra)\nonumber\\
		&&\!\!\!\!\!\!\!\!\!\! =\Tr\Big(\big(\bY-\hat\bH\hat\bX\big)\tra\big(\bY-\hat\bH\hat\bX\big)\Big)
		+\Tr\Big(\hat\bX\hat\bX\tra\Tr(\bU_H)\bV_H\nonumber\\
		&&+\Tr(\bV_X)\bU_X\hat\bH\tra\hat\bH+\Tr(\bV_X)\bU_X\Tr(\bU_H)\bV_H\Big).
	\end{eqnarray}
	By simplifying the above result, we obtain \eqref{eq:ComputeC}.

\end{appendices}	

\bibliographystyle{IEEEtran}
\bibliography{bibliography}

\end{document}